\documentclass[usenatbib]{mn2e}
\usepackage{amsmath}
\usepackage{natbib}
\usepackage{epsfig}
\usepackage{txfonts}
\usepackage{color}

\newcommand{\zmu}{{z_{\mu}}}

\newcommand{\taudot}{\dot{\tau}}
\newcommand{\kD}{k_{\rm D}}
\newcommand{\rs}{r_{\rm s}}
\newcommand{\cs}{c_{\rm s}}

\newcommand{\id}{{\,\rm d}}

\newcommand{\beq}{\begin{equation}}   %

\newcommand{\eeq}{\end{equation}}   %

\newcommand{\beqa}{\begin{eqnarray}}   %

\newcommand{\eeqa}{\end{eqnarray}}   %

\newcommand{\beal}{\begin{align}}

\newcommand{\enal}{\end{align}}

\newcommand{\bspl}{\begin{split}}

\newcommand{\espl}{\end{split}}

\newcommand{\bsub}{\begin{subequations}}

\newcommand{\esub}{\end{subequations}}

\newcommand{\bmulti}{\begin{multline}}   %

\newcommand{\beqm}{\begin{mathletters}}   %

\newcommand{\eeqm}{\end{mathletters}}   %

\newcommand{\Ne}{N_{\rm e}}

\newcommand{\sigT}{\sigma_{\rm T}}

\newcommand{\vek} [1]{{\mathbfit #1}}

\newcommand{\pot}[2]{#1 \times 10^{#2}}

\newcommand{\Yp}{Y_{\rm p}}

\newcommand{\expf}[1]{{{\rm e}^{#1}}}

\newcommand{\zmuy}{{z_{\mu,y}}}

\newcommand{\Mpc}{{\rm Mpc}}

\newcommand{\lsim}{\lesssim}
\newcommand{\gsim}{\gtrsim}

\usepackage{hyperref}
\usepackage{grffile}
\usepackage{graphics}
\usepackage{booktabs}

\title[Isocurvature modes]
{CMB spectral distortions from small-scale isocurvature fluctuations}

\author[Chluba and Grin]{J.~Chluba$^{1, 2}$\thanks{E-mail:
  jchluba@pha.jhu.edu} and D.~Grin$^{3}$\thanks{E-mail:
  dgrin@ias.edu} 
\\
$^{1}$ Department of Physics and Astronomy, Johns Hopkins University, Bloomberg Center 435, 
3400 N. Charles St., Baltimore, MD 21218, USA
  \\
$^{2}$ Canadian Institute for Theoretical Astrophysics, 60 St. George Street,
Toronto, ON M5S 3H8, Canada
\\
$^{3}$ School of Natural Sciences, Institute for Advanced
     Study, Princeton, NJ 08540, USA}

\begin{document}

\date{{Accepted 2013 June 18. Received 2013 April 16}}

\maketitle

\begin{abstract}
The damping of primordial perturbations at small scales gives rise to distortions of the cosmic microwave background (CMB). Here, the dependence of the distortion on the different types of cosmological initial conditions is explored, covering adiabatic, baryon/cold dark matter isocurvature, neutrino density/velocity isocurvature modes and some mixtures. The radiation transfer functions for each mode are determined and then used to compute the dissipative heating rates and spectral distortion signatures, utilizing both analytic estimates and numerical results from the thermalization code \textsc{CosmoTherm}.
Along the way, the early-time super-horizon behavior for the resulting fluid modes is derived in conformal Newtonian gauge, and tight-coupling transfer function approximations are given. CMB spectral distortions caused by different perturbation modes can be estimated using simple $k$-space window functions which are provided here. 
Neutrinos carry away some fraction of the primordial perturbation power, introducing an overall efficiency factor that depends on the perturbation type. 
It is shown that future measurements of the CMB frequency spectrum have the potential to probe different perturbation modes at very small scales (corresponding to wavenumbers $1\,\Mpc^{-1}\lesssim k \lesssim \pot{\rm few}{4}\,\Mpc^{-1}$). These constraints are complementary to those obtained at large scales and hence provide an exciting new window to early-universe physics.
\end{abstract}


\section{Introduction}
\label{sec:Intro}

It is well-known that energy release in the early Universe leads to spectral distortions (SDs) of the CMB \citep{Zeldovich1969,Sunyaev1970mu,Illarionov1975, Illarionov1975b,Danese1982, Burigana1991, Hu1993}. 
At early times, a $\mu$-type distortion is created, while energy release at lower redshifts ($z\lesssim \pot{5}{4}$) results in a $y$-type distortion, similar to the Sunyaev-Zeldovich effect from galaxy clusters.
Constraints on the $\mu$- and $y$-parameters obtained with {\it COBE}/FIRAS \citep{Mather1994, Fixsen1996} limit possible deviations from a blackbody to $\mu\lesssim\pot{9}{-5}$ and $y \lesssim\pot{1.5}{-5}$ at 95\% confidence \citep{Fixsen1996}. At slightly lower frequencies, there are similar limits to $\mu$ from the {\it ARCADE} \citep{Kogut2004, Kogut2006ARCADE, arcade2} and {\it TRIS} \citep{tris1, tris2} experiments. 

Improvements in experimental design and technology may soon allow much more sensitive (a factor of $\simeq 10^{3}-10^4$ in $\mu$ beyond the FIRAS limit!) measurements of the CMB frequency spectrum, as suggested for the proposed experiment {\it PIXIE} \citep{Kogut2011PIXIE}. This has spurred renewed theoretical interest in the cosmological thermalization problem \citep[e.g.,][]{Chluba2011therm, Pajer2012, Khatri2012b} and the use of SDs to probe new physics. For instance, energy injection from dark matter annihilation or decay at redshifts $z\lsim \pot{\text{few}}{6}$ \citep{Hu1993b, McDonald2001, Chluba2010a, Chluba2011therm}, cosmic strings \citep{Ostriker1987, Tashiro2012, Tashiro2012b}, primordial magnetic fields \citep{Jedamzik2000}, but also more exotic possibilities \citep{Lochan2012, Bull2013} could produce a detectable SD signal.

It has long been known that Silk damping \citep{Silk1968} of primordial small-scale perturbations also causes energy release in the early universe \citep{Sunyaev1970diss,Daly1991,Barrow1991,Hu1994}. 
The SD signal depends on the \textit{amplitude} and \textit{shape} of the primordial fluctuation power spectrum, which in turn depend sensitively on the early-universe physics seeding these perturbations. The detection and characterization of CMB SDs could thus offer a powerful new probe of inflationary models \citep{Chluba2011therm, Khatri2011BE, Pajer2012, Dent2012, Ganc2012, Chluba2012inflaton,Powell2012}.
The proper interpretation of measurements of the CMB frequency spectrum in the context of early universe physics requires an accurate treatment of acoustic mode dissipation \citep[as developed in][]{Chluba2012, Khatri2012short2x2, Pajer2012b}, which we follow here. 

The amplitude of the distortion also depends on the {\it type} of perturbation modes \citep{Barrow1991,Hu1994isocurv, Dent2012}. The combined system of fluid + Einstein equations for cold dark matter (CDM), baryons, neutrinos, and photons\footnote{In the following, for associated variables we use subscripts $\rm c$, $\rm b$, $\nu$ and $\gamma$, respectively.} has a variety of propagating (and growing) normal modes. 

Empirically, the most prominent is the \textit{adiabatic} (AD) mode, for which the curvature perturbation $\zeta$ is non-zero, but all initial entropy fluctuations of other species relative to photons vanish:
$S_{i\gamma}=\delta n_{i}/n_{i}-\delta n_{\gamma}/n_{\gamma}
=\delta \rho_{i}/[\left(1+w_{i}\right)\rho_{i}]-(3/4)\, \delta \rho_{\gamma}/\rho_{\gamma}\equiv 0$.
Here, $w_{i}$ denotes the equation of state of the $i^{\rm th}$ species. 
The other well-behaved fundamental modes (on super-horizon scales) are known as \textit{isocurvature} or \textit{entropy} fluctuations, and are characterized by initial values $\zeta=0$ and $S_{i\gamma}\neq 0$. 
They may be simply thought of as spatial fluctuations in the \textit{composition} (or its time derivative) of the Universe \citep[e.g., see][for additional details on perturbation equations and definitions]{Ma1995, Bucher2000}.
The most straightforward isocurvature modes are density isocurvature fluctuations, which naturally sort into the CDM isocurvature (CI), the baryon isocurvature (BI), and the neutrino density isocurvature  (NDI) mode. 
Additionally, there is a neutrino velocity isocurvature mode (NVI), with initial $\zeta=\delta \rho=0$, but a relative velocity $\varv_{\nu}-\varv_{\gamma}$ perturbation between photons and neutrinos.
 
The standard lore is that isocurvature fluctuations behave differently from adiabatic perturbations, sourcing a much larger Sachs-Wolfe effect at large angular scales in the CMB, and acoustic peak phases  that are out of phase with those produced by adiabatic fluctuations \citep{Hu1995CMBanalytic, Hu1996anasmall,Kodama1986, Efstathiou1986, Efstathiou1987}. Indeed, this is true of BI and CI perturbations. NDI modes, on the other hand, behave more like adiabatic modes, since the initially perturbed species (neutrinos) is relativistic, and the isocurvature condition $(\delta \rho=0$) requires an initial energy density perturbation in the photons $(\delta \rho_{\gamma}=-\delta \rho_{\nu})$. 
Similarly, for the NVI a significant CMB dipole is present.
The phase structure of the CMB anisotropy transfer functions for NDI and NVI perturbations thus bears a closer resemblance to the adiabatic transfer function than do the transfer functions for BI/CI modes \citep[for examples, see][]{Kawasaki2012}.

At large scales, corresponding to $10^{-4}\,\Mpc^{-1}\lesssim k\lesssim 1\,\Mpc^{-1}$, these important differences are readily probed by precise measurements of CMB anisotropies, and are highly constrained by the \textsc{Boomerang} \citep{2006ApJ...647..823J} and WMAP \citep{Larson2011} experiments, as well as large-scale structure (LSS) measurements \citep{Beltran2005, Seljak2006, Zunckel2011, Kasanda2012} and other CMB experiments \citep{Moodley2004, MacTavish2006, Bean2006, Dunkley2009}. 
We thus know that at  {\it large scales} the primordial fluctuations are predominantly adiabatic, and there is no evidence of a significant isocurvature component to the primordial initial conditions.  
Recent results from the Planck collaboration \citep{Planck2013iso} strengthen this view, although the hemispherical power asymmetry \citep{Eriksen2004, Planck2013power} could be interpreted as a hint for a modulated large-scale isocurvature mode \citep{Dai2013}.

The implications of this fact are dramatic. In the simplest inflationary models, fluctuations in all species are seeded by quantum fluctuations of a single scalar field; the consistency of observations with adiabatic initial conditions supports these scenarios \citep{bardeen, guthpi,lyth,maldacena2002}. Indeed, the \textsc{Boomerang}, WMAP and Planck data impose very strong constraints on topological defect-dominated models for cosmic structure formation \citep{Contaldi1999, Albrecht2000, Fraisse2005, Planck2013topo}.
Therefore, a detection of a sub-dominant but non-zero isocurvature component would imply deviations from the simplest inflationary picture. For example, dark matter could be composed of  axions \citep{sikivie2008,cademuro}. The axion field would be present and energetically sub-dominant during inflation, exciting the CI mode \citep{axenides1983,linde1985,seckel1985,turner1991}. Isocurvature constraints thus limit the parameter space available for axion dark matter, and SDs might help shed light on this.

Alternatively, in the curvaton model, a sub-dominant scalar field (the \textit{curvaton}) picks up quantum fluctuations, comes to dominate the cosmic energy budget and then seeds a correlated mixture of adiabatic and isocurvature fluctuations. The amplitudes of CI, BI and NDI fluctuations are then set by the relative placement of the epochs of dark matter production, lepton number creation, or baryon number creation, relative to the time of curvaton decay \citep{2002PhLB..524....5L,2003PhRvD..67b3503L,Gordon2003,Gordon2009}. It would be useful to determine if SDs offer any additional leverage on curvaton parameter space.

The curvaton model may also excite {\it compensated isocurvature perturbations} (CIPs), for which $\delta \rho=\zeta=0$, and $\delta S\simeq \delta n_{\rm c}/n_{\rm c}-\delta n_{\rm b}/n_{\rm b}\neq 0$. Surprisingly, current CMB data analyses do not impose constraints to CIPs, but they could soon be detected using higher order correlations of the CMB \citep{2010ApJ...716..907H,2011PhRvD..84l3003G,2011PhRvL.107z1301G}. It would be interesting to see if SDs could be used to detect CIPs or other curvaton-induced modes on small scales, a possibility analyzed in this work. We find that SDs are only useful for this purpose at very futuristic sensitivity levels.

Another interesting theoretical possibility is inhomogeneous baryogenesis, which predicts the existence of BI modes \citep{peebles_iso_a,peebles_iso_b,1999PhRvL..82.2632K}. More broadly, multi-field inflationary models excite isocurvature fluctuations \citep{2001astro.ph.12523G,2001PhRvD..63b3506G}. Constraints to all these possibilities from CMB data are informative, and promise to be even more sensitive with the next \textit{Planck} cosmology data release, including the full temperature and polarization information. 

CMB SDs could allow the characterization of primordial fluctuations on length-scales {\it far smaller} than possible with CMB anisotropy measurements. Most recent work on CMB SDs explores their dependence on the power spectrum of primordial fluctuations, restricting attention to the adiabatic mode \citep{Chluba2012, Chluba2012inflaton, Powell2012, Khatri2013forecast}. SDs could, however, also test for the presence of isocurvature initial conditions, providing a complementary probe to measurements of CMB anisotropies and cosmological large-scale structure on {\it radically} different length-scales than those measurements.

The imprint of isocurvature fluctuations on the CMB spectrum was first studied in detail by \citet{Hu1994isocurv}, where BI models were tightly constrained. In BI models, stars would form early, leading to early reionization and a detectable Compton $y$-type distortion of the CMB frequency spectrum. More recently, the early energy release from BI/CI fluctuations was investigated by \citet{Dent2012}, using a simplified treatment of both the transfer function that maps primordial fluctuations to moments of the radiation field, and of the heating rate itself. 

Recent work has shown the heating process is only $3/4$ as efficient as previous estimates, and that closer to recombination, baryon loading and second-order Doppler terms are important to the heating rate calculation \citep{Chluba2012}. Furthermore, free-streaming relativistic particles, like neutrinos, carry away some fraction of the perturbation power, introducing a dependence on the effective number of relativistic species.
These details are carefully considered here but were omitted in previous work.

Here, we revisit the problem, using the perturbation and heating modules of the SD code {\sc CosmoTherm}\footnote{{\sc CosmoTherm} is available at \url{www.Chluba.de/CosmoTherm}.} \citep{Chluba2011therm, Chluba2012} to precisely calculate both the evolution of fluid and radiation variables from isocurvature initial conditions, as well as the resulting effective plasma heating rate and CMB spectral distortion signal. We compute the SD signal from BI and CI models, improving the estimates of \citet{Dent2012}, and extend our reach to the SD signature of NDI and NVI modes, as well as the CIP (baryon-CDM isocurvature) mode. 
We expand on \citet{Dent2012} to compute both $\mu$ and $y$-type SDs for adiabatic and isocurvature modes, exploring the dependence of the signal on the spectral index of the initial power spectrum of each mode. We compare these different possibilities with the sensitivity of the proposed {\it PIXIE} mission and existing limits to the chemical potential $\mu$ and Compton $y$-parameter from {\it COBE}/FIRAS. We also provide simple analytic expressions for the heating rate of different perturbation types, as well as $k$-space window functions which may be used to estimate the SD ($\mu$ and $y$) signal of arbitrary power spectra.

We begin in Section \ref{sec:distortions} with a review of acoustic mode dissipation and the resulting plasma heating responsible for SDs. In Section \ref{sec:C_i}, we move on to discuss the different families of cosmic initial conditions, and the resulting acoustic mode amplitudes at small scales. We also consider mode mixtures in this section, but restrict ourselves to the simplest cases, providing a simple recipe for correlated, uncorrelated and anticorrelated modes.
In Section \ref{sec:heating_rates}, we present the precise numerical heating rates produced by different acoustic modes. In Section \ref{modelsec}, we discuss possible constraints to the power spectra of different pure modes and some representative mixtures of different modes, as well as implications of SD experiments for specific early-universe scenarios, such as the curvaton model. We conclude in Section \ref{sec:conclusions}.

\section{CMB spectral distortions caused by the dissipation of acoustic modes}
\label{sec:distortions}
In this section, we briefly review how the dissipation of acoustic modes creates spectral distortions. For a more in-depth discussion, see \citet{Chluba2012} and \citet{Khatri2012short2x2}. 
The problem boils down to computation of the effective heating rates for different perturbation modes.
These are obtained both numerically, by solving the cosmological perturbation equations \citep[see][for details]{Ma1995} with {\sc CosmoTherm}, and analytically in the tight-coupling approximation. 
With the heating rates in hand, we use simple analytic estimates to compute the resulting chemical potential, $\mu$, and Compton $y$-parameter.
Generally, the {\it detailed} shape of the distortion is not just represented by a simple superposition of $\mu$- and $y$-distortion, as shown in \citet[][e.g., see Figs. 15 and 19]{Chluba2011therm} and more recently by \citet{Khatri2012mix} or \citet{Chluba2013Green}. These details will, however, be addressed in some future work, since for estimates the approach presented here suffices.

\subsection{Estimates for the $\mu$- and $y$- parameters caused by early energy release}
At high redshifts, $z\gg\zmu\approx \pot{1.98}{6}$, the thermalization process is extremely efficient. As a result, any energy release just increases the specific entropy of the Universe, and thus raises the average temperature of the CMB without producing SDs. For lower redshifts, $z\lesssim \zmu$, thermalization becomes less efficient, and energy release can produce SDs.
For all $z\ll \zmu$, energy injection \textit{initially} appears as a $y$-distortion to the CMB blackbody. If additionally, $z\gg \zmuy\approx \pot{5}{4}$, Comptonization of the radiation field is still efficient and the initial $y$-distortion is mostly converted into a chemical potential $\mu$. On the other hand, if $z\ll \zmuy$, Comptonization is inefficient, and the SD take the form of a non-zero Compton $y$-parameter, and essentially amounts to an early-universe analogue to the Sunyaev-Zeldovich effect.
Thus, a $\mu$-distortion is created by energy release at $\pot{5}{4}\lesssim z\lesssim \pot{2}{6}$ and a $y$-distortion at $z\lesssim \pot{5}{4}$ \citep[see][for more details]{Hu1993}.

To estimate the values of the chemical potential, $\mu$, and Compton $y$-parameter, it is sufficient to compute the effective energy release during the corresponding epochs.
Defining the {\it distortion visibility function}, $\mathcal{J}_{\rm bb}(z)\approx\exp\left(-[z/\zmu]^{5/2}\right)$, the weighted total energy release in the $\mu$- and $y$-era is
\bsub
\label{eq:def_mu_y_Dr_r}
\beal
\label{eq:def_mu_y_Dr_r_a}
\left.\frac{\Delta\rho_\gamma}{\rho_\gamma}\right|_{\mu}
&\approx 
\int_\zmuy^\infty  \frac{\mathcal{J}_{\rm bb}(z)}{a^4 \rho_\gamma}\,
\frac{\id (a^4 Q_{\rm ac})}{\id z} \id z
\\[1mm]
\label{eq:def_mu_y_Dr_r_b}
\left.\frac{\Delta\rho_\gamma}{\rho_\gamma}\right|_{y}
&\approx 
\int^\zmuy_0 \frac{1}{a^4 \rho_\gamma}\,\frac{\id (a^4 Q_{\rm ac})}{\id z} \id z.
\end{align}
\esub
We introduced the energy release caused by the dissipation of primordial acoustic modes\footnote{Alternatively, one can write $a^{-4} \rho^{-1}_\gamma \,\id (a^4 Q_{\rm ac})/\id z\approx \id (Q_{\rm ac}/\rho_\gamma)/\id z$.}, $a^{-4} \rho^{-1}_\gamma \,\id (a^4 Q_{\rm ac})/\id z$; however, any process leading to energy release can be added here.
The factors of the scale factor (normalized to unity today) $a=(1+z)^{-1}$ cancel the main redshift dependence of the background radiation field, which is irrelevant for the creation of SDs. 
The factor $\mathcal{J}_{\rm bb}(z)$ parametrizes the thermalization efficiency accounting for the effects of photon production/destruction by double Compton scattering \citep[see][for more details]{Danese1982, Burigana1991, Hu1993, Chluba2011therm, Khatri2012b}.

With the simple expressions from 
\citet{Sunyaev1970mu}, 
$\mu \approx 1.4 \,\Delta \rho_\gamma/\rho_\gamma|_\mu$ and $y \approx \frac{1}{4} \Delta \rho_\gamma/\rho_\gamma|_y$, Eq.~\eqref{eq:def_mu_y_Dr_r} can be used to estimate the expected distortion at high frequencies.
This imposes upper limits to any energy-releasing process in the early Universe. The relative ratio of the $\mu$- and $y$-parameters in principle can be further used to distinguish different sources of early energy release via their redshift dependence, although this is possible only for specific models of the thermal history.
At low redshifts, after the recombination epoch, many mechanisms [e.g., reionization \citep{Hu1994pert}; supernova heating \citep{Oh2003}; large-scale structure formation shocks \citep{Sunyaev1972b, Cen1999, Miniati2000}; unresolved Sunyaev-Zeldovich clusters and the warm-hot intergalactic medium \citep{Markevitch1991, daSilva2000, Zhang2004}] give rise to large average $y$-type distortions.
As a result, early-universe processes which generate a $\mu$-type SD are more readily constrained with measurements of the CMB frequency spectrum, although useful bounds may still be derived from measurements of or limits on $y$-type distortions.
One way to distinguish $y$-distortions from the pre-recombination epoch from those created at later stages might be the cosmological recombination radiation \citep{Chluba2006b, Chluba2008c, Sunyaev2009, Chluba2010a}; also spectral-spatial information could be used to disentangle different sources of $y$-distortions, but a more detailed discussion is beyond the scope of this paper.
To make further progress, we must specify the effective heating rate, $Q_{\rm ac}$, and its dependence on the cosmological initial conditions.

\subsection{The effective heating rate from acoustic damping}
\label{sec:heating_rate}
Small-scale perturbations of the photon temperature are completely erased by shear viscosity and thermal conduction \citep{WeinbergBook}. 
These processes isotropize the photon-baryon fluid and lead to the mixing of blackbodies with slightly different temperatures \citep{Zeldovich1972, Chluba2004} causing an increase of the local average photon temperature and a $y$-type SD.
The spatially averaged SD source function, $\left<\mathcal{S}_{\rm ac}\right>$, directly depends on the amplitude and shape of the primordial perturbation power spectrum, $P_i(k)$, as well as the detailed evolution of moments of the radiation field for given initial conditions.
It is determined by \citep{Chluba2012, Khatri2012short2x2}:
\begin{align}
\label{eq:Sac_full}
\left<\mathcal{S}_{\rm ac}\right>
=&
\int \frac{k^2\id
  k}{2\pi^2}P_i(k)
  \left[\frac{\left(3\Theta_1-\varv\right)^2}{3}+\frac{9}{2}\Theta_2^2
  \nonumber\right.
  \\
  &\qquad\qquad
  \left.
  -\frac{1}{2}\Theta_2\left(\Theta_2^{\rm P}+\Theta_0^{\rm P}\right)+\sum_{\ell\ge 3}(2\ell+1)\Theta_{\ell}^2\right],
\end{align}
where $\Theta_\ell$ and $\Theta^{\rm P}_\ell$ denote the photon temperature and polarization transfer functions and $\varv$ the one for the baryon velocity. 
This source function can be computed accurately using the cosmological thermalization code {\sc CosmoTherm}.

Given $\left<\mathcal{S}_{\rm ac}\right>$, the required effective energy release rate caused by the damping of acoustic modes is determined by
%
\beal
\label{eq:Q_ac}
\frac{1}{a^4 \rho_\gamma}\frac{\id (a^4 Q_{\rm ac})}{\id z}
& = 
\frac{4\taudot \left<\mathcal{S}_{\rm ac}\right>}{H (1+z)},
\end{align}
where $\taudot =\sigT\Ne c \approx \pot{4.4}{-21}(1+z)^{3}\,{\rm sec^{-1}}$ denotes the rate of Thomson scattering and $H\approx \pot{2.1}{-20}\,(1+z)^2 {\rm sec^{-1}}$ is the Hubble expansion rate\footnote{The approximations for $\taudot$ and $H$ are only valid at high redshifts, during the radiation-dominated era.}.
The factor of 4 arises because a $y$-distortion causes a change in the photon energy density by $\Delta \rho_\gamma\simeq4\rho_\gamma$. The factor $\taudot$ arises because the source function, $\left<\mathcal{S}_{\rm ac}\right> >0$, is defined with respect to the Thomson-scattering time-scale and the factor $1/[H (1+z)]$ is needed for the conversion to $\!\id z$.

Below we compute the effective heating rate for different initial perturbation modes, with particular focus on AD, BI, CI, NDI, NVI modes (see Sect.~\ref{sec:heating_rates} and Fig.~\ref{fig:heating_comp}), and simple mode mixtures. 
The important differences are caused by the transfer functions and their relation to the initial power spectra, which can be understood using some simple analytic approximations deep into the radiation-dominated era, when photons and electrons are tightly coupled.

\subsection{Source term before recombination}
In this work we are particularly interested in energy release well before the recombination epoch ($z\gtrsim 10^4$). 
At that time, the Universe is still radiation-dominated and with small baryon loading $R=3\rho_{\rm b}/4\rho_\gamma\approx 673 \, (1+z)^{-1}\lesssim 7\%$.
Also, photons and baryons are tightly coupled so that $(3\Theta_1-\varv)\simeq 0$. Furthermore, higher order temperature perturbations with $\ell>2$ are negligible (Thomson scattering isotropizes the radiation field) and the dissipation physics is mainly determined by the quadrupole anisotropy.
In this limit\footnote{Note that in our definition $\Theta^{\rm Hu}_\ell =(2\ell+1)\Theta_\ell$.}, $\taudot\,\Theta_2\simeq \frac{8}{15} k \Theta_1$ and $\Theta_2^{\rm P}+\Theta_0^{\rm P}\simeq \frac{3}{2}\,\Theta_2$ \citep{Hu1996anasmall}, so that
\begin{align}
\label{eq:Sac_tc}
\left<\mathcal{S}_{\rm ac}\right>
&\approx
\int \frac{k^2\id
  k}{2\pi^2}P_i(k) \frac{15}{4}\Theta_2^2
  \approx
\frac{1}{\taudot^2} \int \frac{\id
  k}{2\pi^2} k^4 P_i(k) \frac{16}{15}\Theta_1^2.
\end{align}
This result shows that an approximation for the source term can be obtained using analytic expressions for the CMB dipole transfer function. 
Inside the horizon, $\Theta_0$ and $\Theta_1\simeq - \partial_\eta \Theta_0/k$ have the generic form \citep{Hu1996anasmall}
\bsub
\label{eq:theta1_tc}
\begin{align}
\label{eq:theta1_tc_a}
\Theta_0
&\approx
\frac{1}{(1+R)^{1/4}} \left[ A(k)\cos(k\rs) + B(k) \sin(k\rs)\right]\, \expf{-k^2/\kD^2}
\\
\Theta_1
&\approx
\frac{\cs}{(1+R)^{1/4}} \left[ A(k) \sin(k\rs) - B(k)\cos(k\rs)\right]\, \expf{-k^2/\kD^2}
\end{align}
\esub
where $(\cs/c)^2=1/[3(1+R)] \approx 1/3$ is the photon-baryon sound speed, 
$\eta=\int c \id t / a$ denotes conformal time
and the damping scale, $\kD$, is determined by
\beal
\label{eq:k_diss}
\partial_t \kD^{-2}=
\frac{\cs^2}{2 \taudot}\left[\frac{R^2}{1+R}+\frac{16}{15}\right]
\stackrel{\stackrel{R\simeq0}{\downarrow}}{\approx} 
\!\frac{8}{45 \taudot}.
\end{align}
Furthermore, $\rs$ is the sound horizon, and the WKB amplitudes $A(k)$ and $B(k)$ are determined by the initial condition (see Sect.~\ref{sec:C_i}).

For a given wavenumber $k$ the source function oscillates rapidly, but for the net effect of many modes on the CMB spectrum we are only interested in  time-averaged values. Squaring Eq.~\eqref{eq:theta1_tc} and averaging over many periods, interference terms vanish and by replacing $\sin^2(k\rs)\rightarrow 1/2$ and $\cos^2(k\rs)\rightarrow 1/2$ we obtain
\begin{align}
\label{eq:Sac_tc_new}
\left<\mathcal{S}_{\rm ac}\right>
&\approx
\frac{8}{45\taudot^2} \int \frac{\!\id
  k}{2\pi^2} k^4 P_i(k) \left[ A^2(k) + B^2(k) \right]\, \expf{-2k^2/\kD^2}
  \nonumber
  \\
  &\approx \frac{1}{\taudot}\partial_t \kD^{-2} 
  \int \frac{ \!\id
  k}{2\pi^2} k^4 P_i(k)  \,C^2(k)\, \expf{-2k^2/\kD^2}
   \nonumber
  \\
  &\approx -\frac{1}{2 \taudot}
  \int \mathcal{P}_i(k) \, C^2(k)\, \partial_t \expf{-2k^2/\kD^2} \id\ln k .
\end{align}
Here, we defined $C^2(k)=A^2(k) + B^2(k)$ and $\mathcal{P}_i(k)=k^3 P_i(k)/(2\pi^2)$.
This expression can be used to estimate the effective heating rate caused by initial fluctuations, characterized by their power spectra and perturbation type (Sec. \ref{sec:heating_rates}).
Furthermore, since a separation of time and scale dependent terms is achieved, it is possible to define $k$-space window functions that can be pre-computed once the cosmology is fixed (Sect.~\ref{sec:estimates}).
Differences between various perturbation modes are then determined by the overall normalization, $C^2(k)$, specifying a mode dependent heating efficiency.

\section{Dependence on the power spectrum and fluid modes at small scales}
\label{sec:C_i}
As equations~\eqref{eq:Sac_full}, \eqref{eq:Q_ac} and \eqref{eq:Sac_tc_new} show, the energy release depends directly on the primordial perturbation power spectrum, $P_i(k)$, which we characterize using the simple parametrization \citep{Kosowsky1995}, 
\beal
\label{eq:P_k_st}
P_i(k)&=2\pi^2 k^{-3} A_i (k/k_0)^{n_i-1+\frac{1}{2} n_{i, \rm run} \ln(k/k_0)}
\end{align}
with amplitude $A_i$, spectral index $n_i$, running $n_{i, \rm run}\equiv \id n_i / \id \ln k$, and pivot scale $k_0$, which we set to $k_0=0.002\,\Mpc^{-1}$. 

The definition of the power spectrum, $P_i(k)$ (in particular, whether or not $P_{i}(k)$ describes density fluctuations or fluctuations of a gauge-invariant variable like $\zeta$ or the entropy $S_{i\gamma}$) may vary, depending on the perturbation type (see below).
For adiabatic perturbations, SD constraints for a variety of theoretically instructive values of $n_{i}$ and $n_{i,\rm run}$ were recently discussed by \citet{Chluba2012, Chluba2012inflaton}, in the context of inflationary theory. Here, we focus on the dependence of the heating rate on different perturbation types, and provide simple expressions for the sub-horizon amplitudes, $C^2(k)=A^2(k)+B^2(k)$, defined by Eq.~\eqref{eq:Sac_tc_new}, of small-scale modes. These allow comparing the heating rates of different perturbation types and also provide simple means for understanding the dependence on the spectral index (Sect.~\ref{sec:heating_rates}). 

To determine the coefficients, $A(k)$ and $B(k)$, of Eq.~\eqref{eq:theta1_tc}, we can resort to analytic approximations \citep[e.g.,][]{Hu1996anasmall}, or simply solve the evolution equations for the fluid and metric variables \citep[see][for definitions]{Ma1995} numerically for different initial conditions (see Appendix~\ref{app:initial} for pure modes) to determine the main dependences on scale and cosmology. The evolution of the potentials and their decay during horizon crossing affect the mode amplitudes in a non-trivial way, and so we use the latter approach to obtain a more accurate but simple description.
Given approximations for $A(k)$ and $B(k)$, the heating rates for general mode mixtures at high redshifts can be constructed. Here, we restrict ourselves to simple mode mixtures, although more general cases with off-diagonal correlations could be of theoretical interest \citep{Moodley2004}.

\begin{figure*}
\centering
\includegraphics[width=1.02\columnwidth]{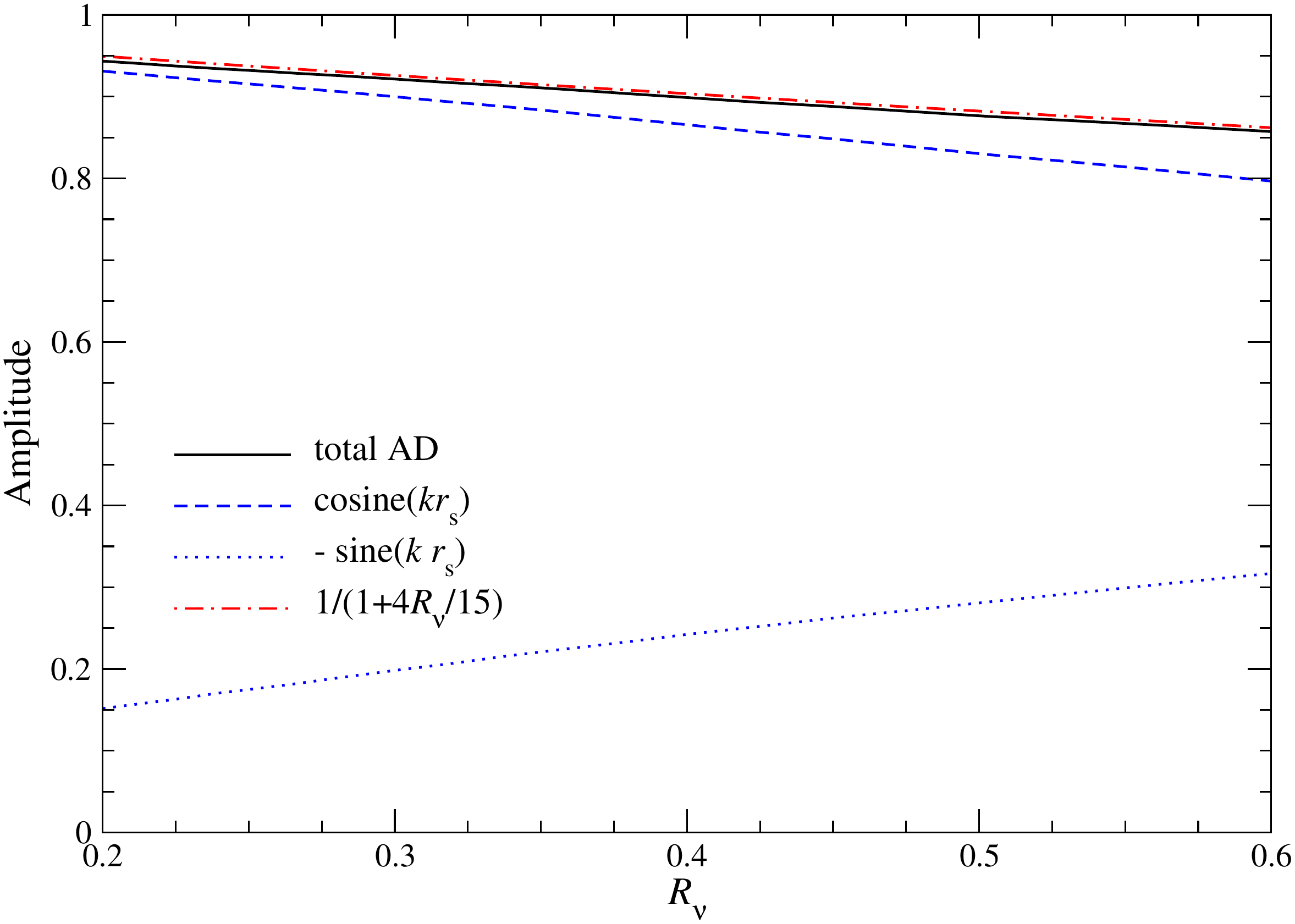}
\hspace{2mm}
\includegraphics[width=1.02\columnwidth]{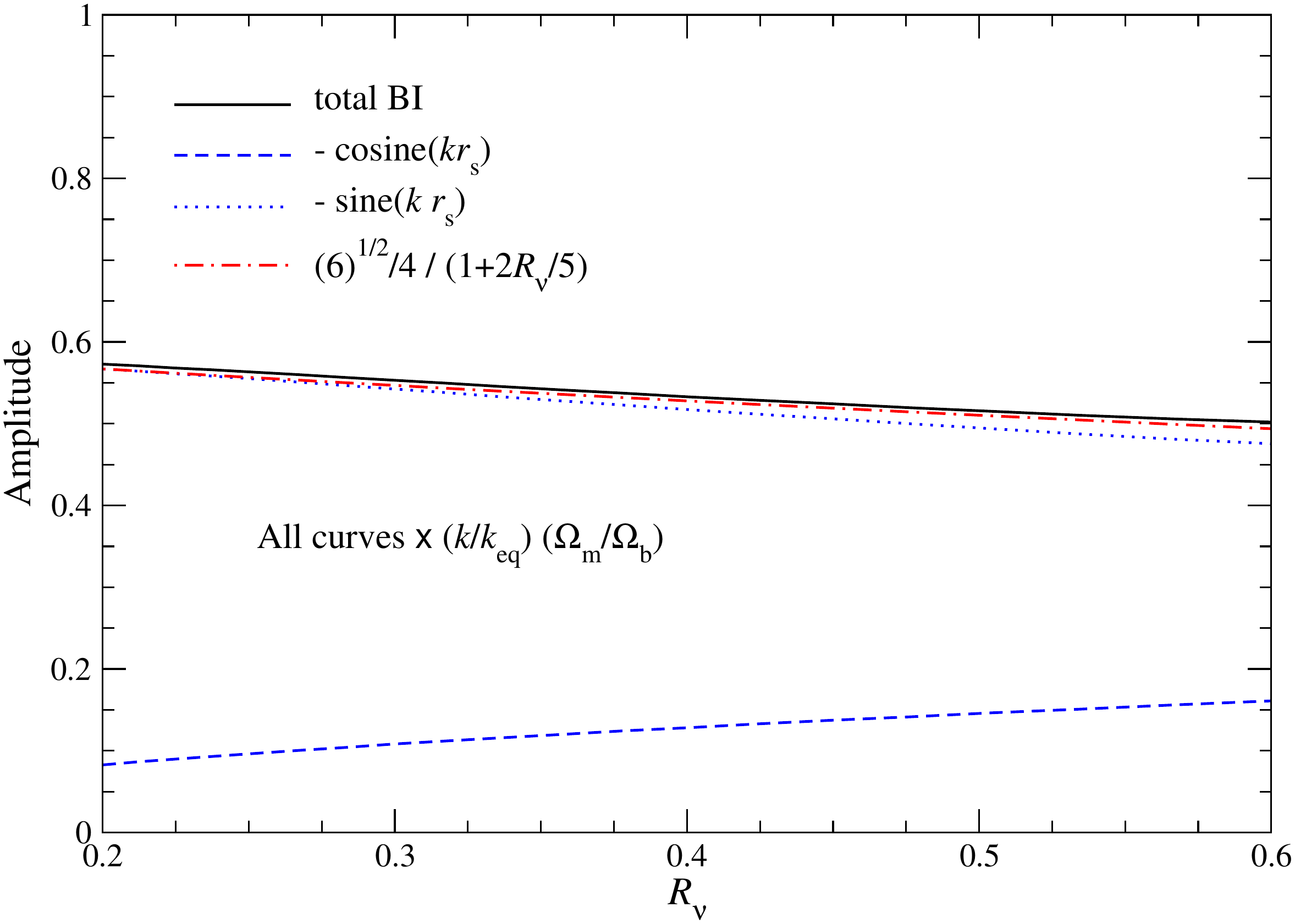}
\\[3mm]
\includegraphics[width=1.02\columnwidth]{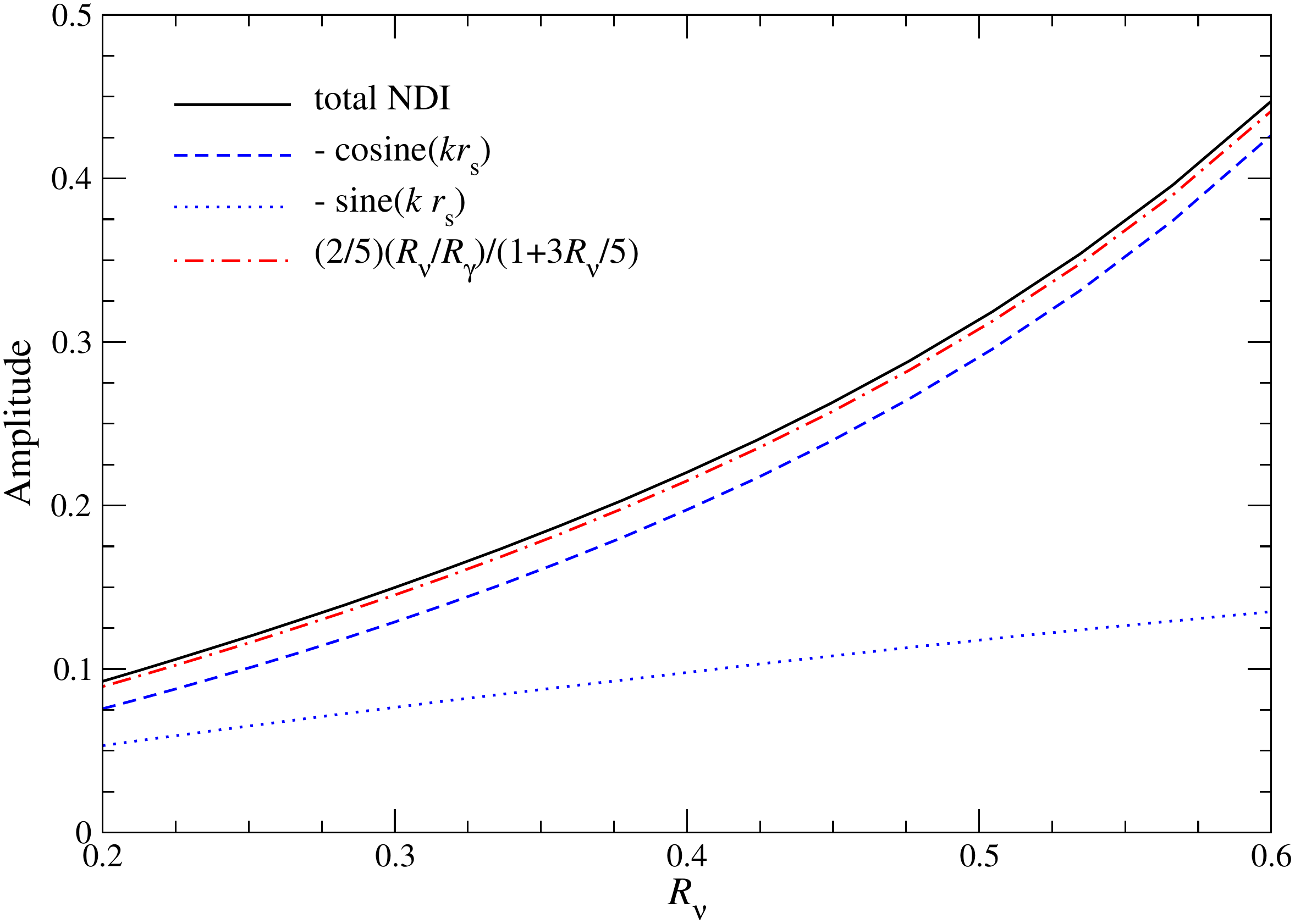}
\hspace{2mm}
\includegraphics[width=1.02\columnwidth]{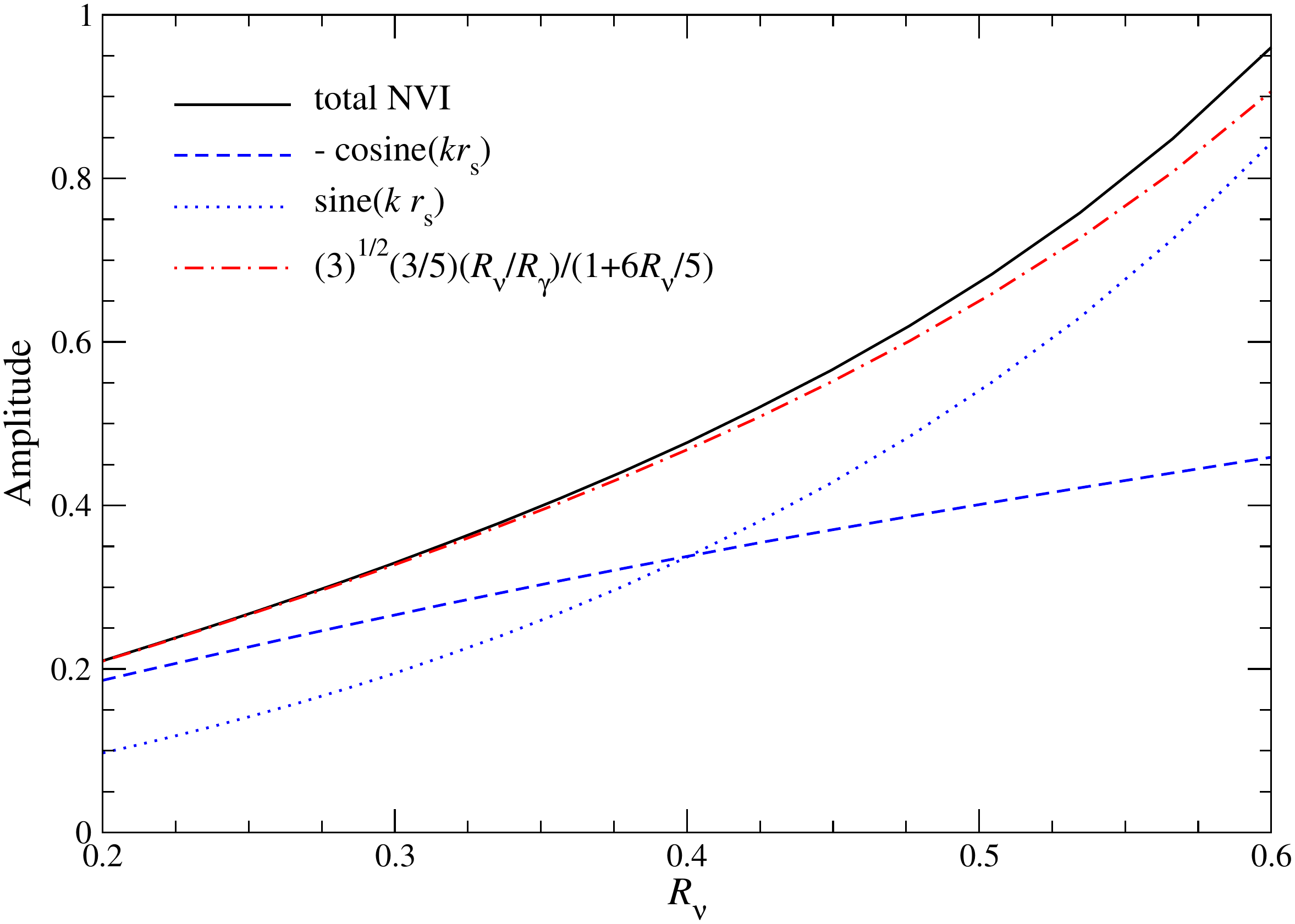}
\caption{Amplitude of cosine and sine parts in the WKB approximation of the sub-horizon monopole transfer functions, Eq.~\eqref{eq:theta1_tc_a}. The CI mode amplitudes are equal to the BI amplitudes times $\Omega_{\rm cdm}/\Omega_{\rm b}$. We numerically solved the fluid and potential equations for $k=10\,\Mpc^{-1}$ as representative example for small-scale modes varying $R_\nu$.
}
\label{fig:Amplitudes}
\end{figure*}

\subsection{Adiabatic mode (AD)}
For adiabatic (isentropic) perturbations, we consider the power spectrum of curvature perturbations, $P_\zeta(k)$.
The WKB mode amplitude at small scales is \citep{Hu1996anasmall} 
\beal
\label{eq:C_ad}
A&\simeq \left(1+\frac{4}{15}R_\nu\right)^{-1}
\end{align}
and $B\simeq 0$. Here, $R_\nu=\rho_\nu/(\rho_\gamma+\rho_\nu)\approx 0.41$ denotes the fractional contribution of massless neutrinos to the energy density of relativistic species, for effective number of relativistic degrees of freedom, $N_{\rm eff}\simeq 3.046$. The term $4R_\nu/15$ accounts for the correction caused by anisotropic stress in the neutrino fluid. It allows neutrinos to carry away some part of the perturbation power, without sourcing any CMB SD. Increasing the effective number of neutrinos therefore decreases the net heating rate and SD. 

Although the initial temperature perturbation of the monopole is about three times smaller than $A$, decay of the potentials after horizon crossing boosts the mode amplitude to this larger value by gravitational forcing. Numerically, we find a small admixture of the sine term, i.e. $B\neq 0$, to the photon monopole transfer function, Eq.~\eqref{eq:theta1_tc_a}, caused by the driving term. As shown in Fig.~\ref{fig:Amplitudes}, the overall amplitude of the small-scale mode is well represented by $|C|\simeq (1+4R_\nu/15)^{-1}$, but 
\beal
\label{eq:C_ad_fit}
A&\simeq 1-0.338 \, R_\nu, & B&\simeq -\pot{7.16}{-2}-0.418\, R_\nu
\end{align}
provide better approximations for the sine and cosine terms. At $R_\nu=0.41$, this gives $A\simeq0.86$, $B\simeq-0.24$ and $C^2\simeq 0.81$. 
The dependence of $A(k)$ and $B(k)$ on $R_\nu$ is illustrated in Fig.~\ref{fig:C_for_mix}.

\subsection{Baryon and CDM isocurvature mode (BI/CI)}
For BI and CI perturbations, the situation is very different. In this case, $P_i(k)$ is defined as the power spectrum of density perturbations, and the WKB mode amplitudes are roughly given by \citep{Hu1996anasmall}
\beal
\label{eq:C_iso}
B&\approx-\frac{\sqrt{6}}{4}\frac{\Omega_{\rm i}}{\Omega_{\rm m}}\,\frac{k_{\rm eq}}{k} 
\left(1-\frac{4}{15}R_\nu\right),
\end{align}
and $A\simeq 0$. Here, $k_{\rm eq}\simeq \pot{9.46}{-2}\Omega_{\rm m} h^2\sqrt{1-R_\nu}\,\Mpc^{-1}\simeq \pot{9.56}{-3}\,\Mpc^{-1}$ is the wavenumber of a mode crossing the horizon at matter-radiation equality, and $\rm i=\{b,c\}$ for baryons and cold dark matter, respectively.
As this expression shows, for scales $k\ll k_{\rm eq}$ the amplitude of photon temperature perturbations is suppressed. This occurs because these modes enter the horizon during radiation domination, when the gravitational sourcing of photon temperature perturbations by baryon/CDM density fluctuations is suppressed. In this case, the dissipative heating rate is larger at late times (during matter domination), when smaller $k$ modes enter the horizon.

From the WKB solution, we can also see that the heating rate decreases as the number of effective neutrino species increases\footnote{For small changes, $\Delta R_\nu/R_\nu\ll 1$, the scaling is similar to that of the adiabatic modes.}. Numerically, we find that the total mode amplitude is represented slightly better by replacing $(1-4R_\nu/15)\rightarrow (1+ 2R_\nu/5)^{-1}$, but the difference is only a few percent. 
For the WKB amplitudes, we find
\bsub
\label{eq:C_BI_CI_fit}
\beal
A&\simeq - \frac{\Omega_{\rm i}}{\Omega_{\rm m}}\,\frac{k_{\rm eq}}{k}\left[\pot{4.79}{-2}+0.195\, R_\nu\right]
\\
B&\simeq - \frac{\Omega_{\rm i}}{\Omega_{\rm m}}\,\frac{k_{\rm eq}}{k}\left[0.613-0.235\, R_\nu \right],
\end{align}
\esub
which gives $A\simeq- 0.13 (\Omega_{\rm i}/\Omega_{\rm m})(k_{\rm eq}/k)$, $B\simeq-0.52(\Omega_{\rm i}/\Omega_{\rm m})(k_{\rm eq}/k)$ and $C^2\simeq 0.28(\Omega_{\rm i}/\Omega_{\rm m})^2(k_{\rm eq}/k)^2$ for $R_\nu=0.41$ (see Fig.~\ref{fig:C_for_mix}).

Our calculations also show that, in contrast to the AD mode, the potentials do not decay as fast after entering the horizon. Consequently, the zero-point of the monopole transfer function is offset by $\simeq \psi$, in agreement with previous analysis \citep[e.g., see][]{Hu1995CMBanalytic,Kawasaki2012}. For the heating rate this aspect is not important, since the local monopole does not source any significant distortion \citep{Chluba2012}.

\subsection{Neutrino density isocurvature mode (NDI)}
Unlike BI/CI modes, NDI modes begin with non-vanishing initial (super-horizon) potential perturbations $\phi$ and $\psi$, which immediately source photon perturbations close to horizon crossing.  Additionally, as a result of the isocurvature condition ($\delta \rho=0$), the initial neutrino density perturbation $\delta \rho_{\nu}$ requires an equal but opposite photon energy density perturbation $\delta \rho_{\gamma}=-\delta \rho_{\nu}$ (see Appendix~\ref{app:initial}). The non-zero photon energy density perturbation means that in principle both the sine and cosine parts of the monopole transfer function are excited. 
Thus, perturbations in the neutrino density immediately start oscillating with appreciable amplitude after horizon crossing, rendering the sine part sub-dominant. From our numerical solutions, we find that at small scales the total amplitude of the neutrino density isocurvature mode is well represented by 
\beal
\label{eq:C_iso_neut}
|C|\simeq \frac{2 R_\nu/R_\gamma}{5(1+3R_\nu/5)},
\end{align}
with $R_\gamma=1-R_\nu\simeq 0.59$.
Since small-scale NDI modes behave similar to AD modes, the SD caused by NDI modes (up to an overall efficiency factor) is also expected to be comparable. 
The cosine and sine amplitudes of the monopole transfer function are
\bsub
\label{eq:C_NDI_fit}
\beal
A&\simeq -(R_\nu/R_\gamma)\left[0.316-\pot{5.25}{-2}\, R_\nu\right]
\\
B&\simeq - (R_\nu/R_\gamma)\left[0.267-0.297\, R_\nu\right],
\end{align}
\esub
which gives $A\simeq- 0.2$, $B\simeq-0.1$ and $C^2\simeq 0.052$ (see Fig.~\ref{fig:C_for_mix}).

\subsection{Neutrino velocity isocurvature mode (NVI)}
NVI modes are excited by initial perturbations in the neutrino velocity, and in this case $P_{i}(k)$ is a power spectrum of the fluid expansion of the photon-neutrino relative velocity, $\theta_{\nu}-\theta_{\gamma}$. Like NDI modes, NVI perturbations immediately start oscillating upon horizon entry, exciting sine and cosine terms comparable with relative amplitudes, depending on $R_\nu$.
In the limit of $R_\nu\ll 1$ mainly the cosine part of $\Theta_0$ is excited, while for $R_\nu\simeq 1$ it is the sine part (see Fig.~\ref{fig:Amplitudes}).
We find that the total amplitude is well approximated by 
\beal
\label{eq:C_iso_neut_velo}
|C| \simeq \frac{3\sqrt{3}\,R_\nu/R_\gamma}{5 (1+6R_\nu/5)}.
\end{align}
The overall heating caused by the NVI mode (like the NDI mode) is thus expected to generate SDs similar to those caused by the AD and NDI mode. 
For the WKB amplitudes, we find
\bsub
\label{eq:C_NVI_fit}
\beal
A&\simeq -(R_\nu/R_\gamma)\left[0.935-1.06\, R_\nu\right]
\\
B&\simeq (R_\nu/R_\gamma)\left[0.349+0.369\, R_\nu\right],
\end{align}
\esub
yielding $A\simeq- 0.35$, $B\simeq0.35$ and $C^2\simeq 0.25$. The heating efficiency for NVI is thus $\simeq 5$ times larger than for the NDI.

\subsection{Mixture of different perturbation modes}
\label{sec:A_mixed}
Assuming that initial perturbations are created by a single field, the different pure modes can be excited in different proportions. With the expressions for $A(k)$ and $B(k)$ given above it is thus straightforward to compute the heating efficiencies. The WKB amplitudes are simply given by coherent superposition
\bsub
\label{eq:C_mix}
\beal
A(k)&\simeq \sum_{i} \alpha_i(k) A_i(k)
\\
B(k)&\simeq \sum_{i} \beta_i(k) B_i(k),
\end{align}
\esub
where the sum runs over AD, BI, CI, NDI, and NVI mode amplitudes. The mixing coefficient, $\alpha_i$ and $\beta_i$ are determined by the initial conditions. This superposition respects the phase of the transfer functions and the effective small-scale mode amplitude and heating efficiency is directly determined by $C^2(k)\simeq A^2(k)+B^2(k)$.
For the special case of totally uncorrelated modes (created by statistically independent processes), one finds
\beal
\label{eq:C_mix_uncorrelated}
C^2_{\rm uncorr}&\simeq \sum_{i} \alpha^2_i(k) A^2_i(k)+\sum_{i} \beta^2_i(k) B^2_i(k)\equiv \sum_i C^2_i,
\end{align}
where destructive and constructive interference terms average out.

\begin{figure}
\centering
\includegraphics[width=1.02\columnwidth]{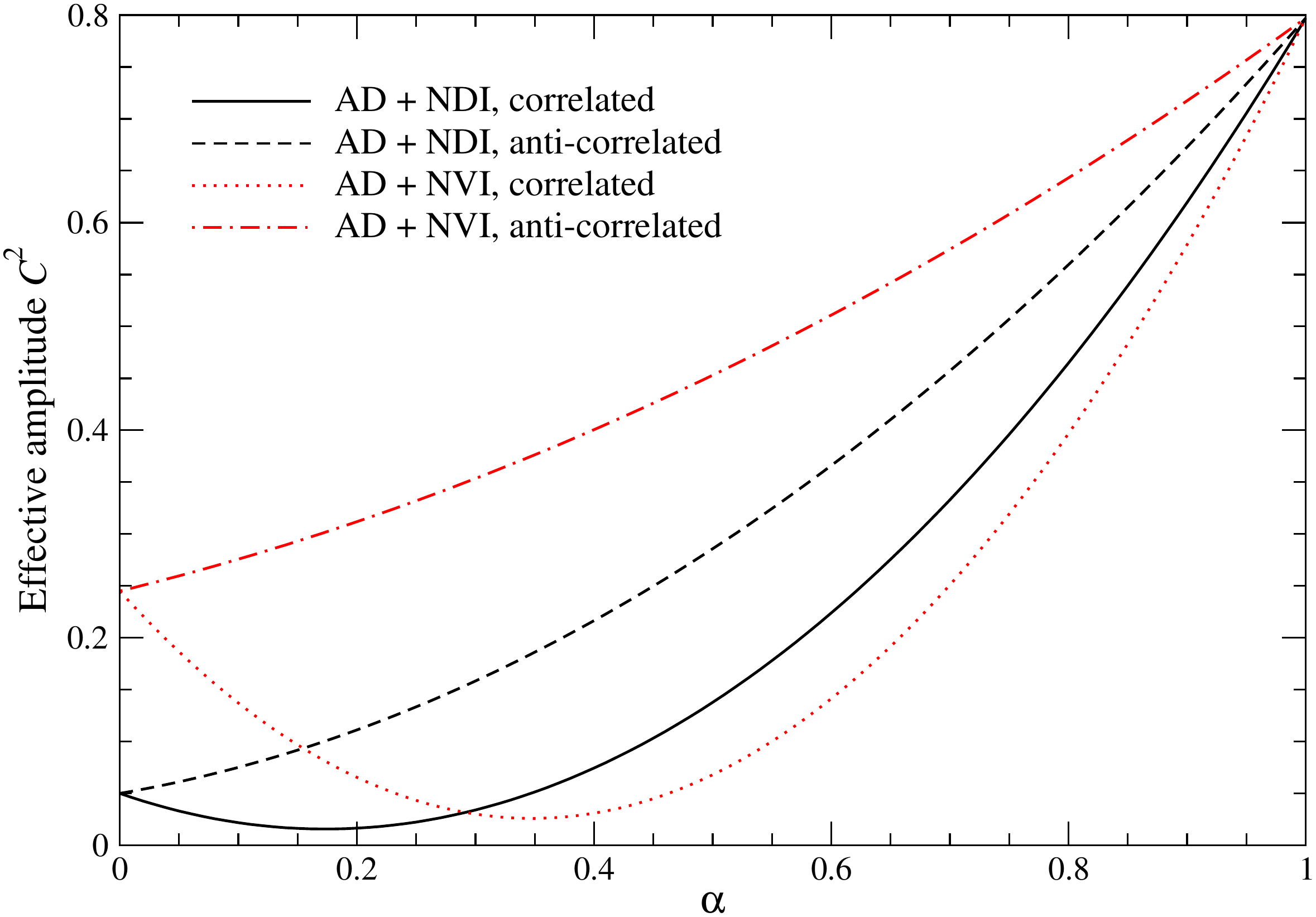}
\caption{
Effective mode amplitude, $C^2$, for mixture of AD with either NDI or NVI modes.
We chose $R_\nu=0.41$ for which we have $A_{\rm AD}=0.86$, $B_{\rm AD}=-0.24$, $A_{\rm NDI}=-0.2$, $B_{\rm NDI}=-0.1$, $A_{\rm NVI}=-0.35$, and $B_{\rm NVI}=0.35$.
}
\label{fig:C_for_mix}
\end{figure}
One simple example for a mode mixture is the CIP, for which (without the loss of generality) the relative ratio of BI and CI amplitude is $\alpha_{\rm CI}=\beta_{\rm CI}=1$ and $\alpha_{\rm BI}=\beta_{\rm BI}=-\Omega_{\rm cdm}/\Omega_{\rm b}$, which gives $A(k)\simeq B(k)\simeq 0$, and thus very small net heating. 
We confirmed this statement numerically, finding that the CIP heating rate is suppressed by at least two orders of magnitude relative to the CI mode, which is itself expected to give a very small distortion (see Sect.~\ref{sec:diss_constr}). We therefore omit CIP modes below.

As an additional simple example, we consider mixtures of AD with either NDI or NVI modes. We can parametrize the relative amplitudes as $\alpha_{\rm AD}=\beta_{\rm AD}=\alpha=\rm const$ and $\alpha_{j}=\beta_{j}=1-\alpha$, where $j\in \{{\rm NDI}, {\rm NVI}\}$ and $\alpha\in [0,1]$. For fully correlated modes, we find:
\bsub
\label{eq:C_mix_correlated}
\beal
A_{\rm corr}&\simeq \alpha A_{\rm AD} + (1-\alpha) A_{j}
\\
B_{\rm corr}&\simeq \alpha B_{\rm AD} + (1-\alpha) B_{j}.
\end{align}
\esub
Similarly, for fully anticorrelated perturbations, we have
\bsub
\label{eq:C_mix_anticorrelated}
\beal
A_{\rm anti}&\simeq \alpha A_{\rm AD} - (1-\alpha) A_{j}
\\
B_{\rm anti}&\simeq \alpha B_{\rm AD} - (1-\alpha) B_{j}.
\end{align}
\esub
These expressions directly determine the effective mode amplitude, $C^2(k)$, which we show in Fig.~\ref{fig:C_for_mix} for $R_\nu=0.41$. For the correlated AD and NDI modes, the net amplitude has a minimum at $\alpha\simeq 0.17$ or an isocurvature-adiabatic ratio of $\left(1-\alpha\right)/\alpha \simeq 4.9$, but due to the phase difference of the AD and NDI transfer functions the amplitude does not vanish completely. Comparing with the heating rate of the uncorrelated mode, this is $C^2_{\rm corr}\simeq (1/4) C^2_{\rm uncorr}$. On the other hand, the anticorrelated mode heating rate is about $C^2_{\rm anti}\simeq 1.7 C^2_{\rm uncorr}$ at $\alpha\simeq 0.2$, with constructive interference dominating.
Similarly, for the AD plus NVI mode, the net amplitude becomes small for $\alpha\simeq 0.35$ or $\left(1-\alpha\right)/\alpha\simeq 1.9$.
Relative to the uncorrelated mode amplitude this is $C^2_{\rm corr}\simeq 0.13 C^2_{\rm uncorr}$. For the AD with NVI anticorrelated mode case, we again find constructive interference with $C^2_{\rm anti}\simeq 1.9 C^2_{\rm uncorr}$ at $\alpha\simeq 0.35$.

The examples above are just meant to illustrate the computation of the effective mode amplitudes, $C^2$, at small scales for different mode mixtures. In a similar way, one can consider AD plus CI or BI modes or even modes with three perturbation types excited. 
A detailed analysis is beyond the scope of this paper.

\begin{figure*}
\centering
\includegraphics[width=1.25\columnwidth]{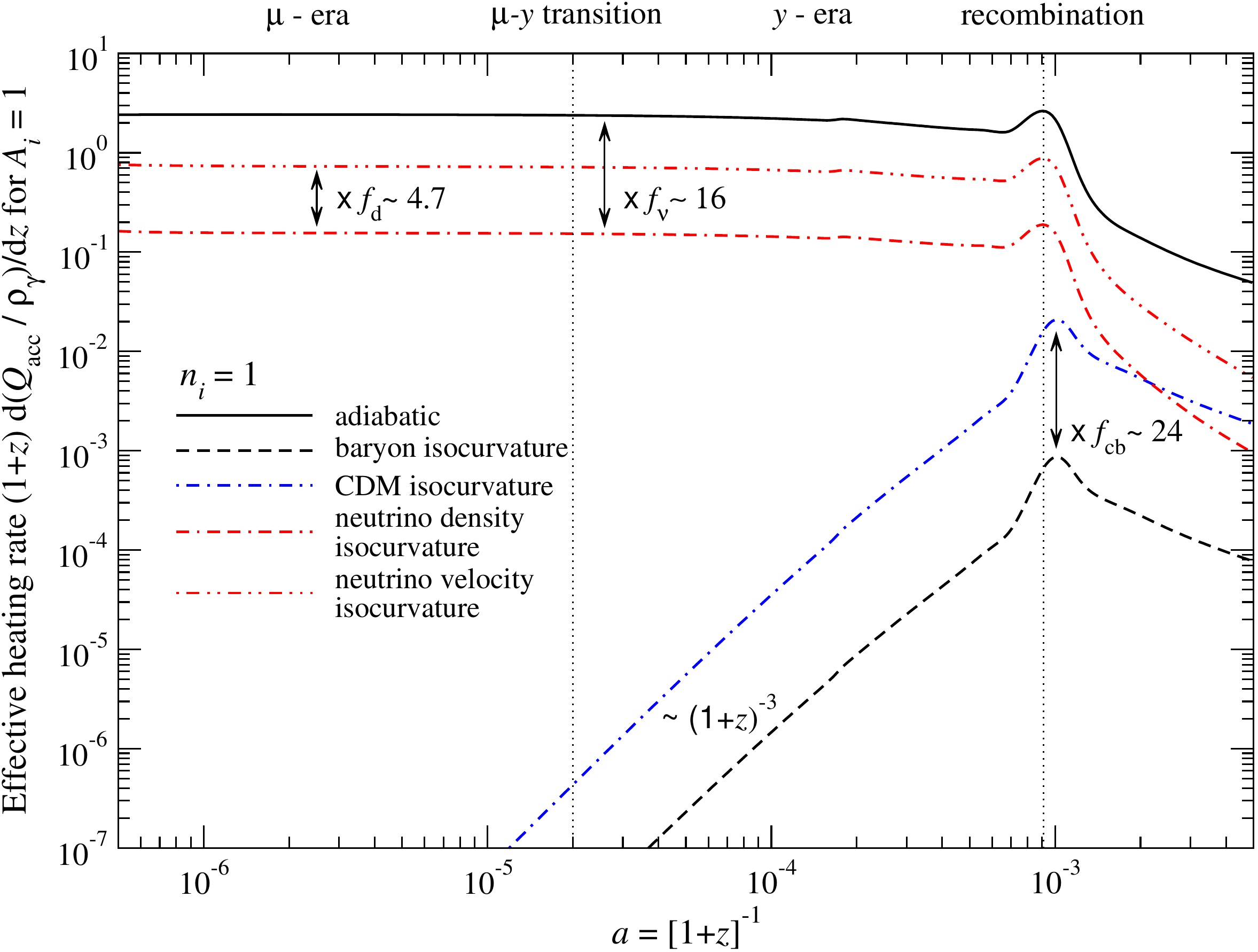}
\\[2mm]
\includegraphics[width=1.25\columnwidth]{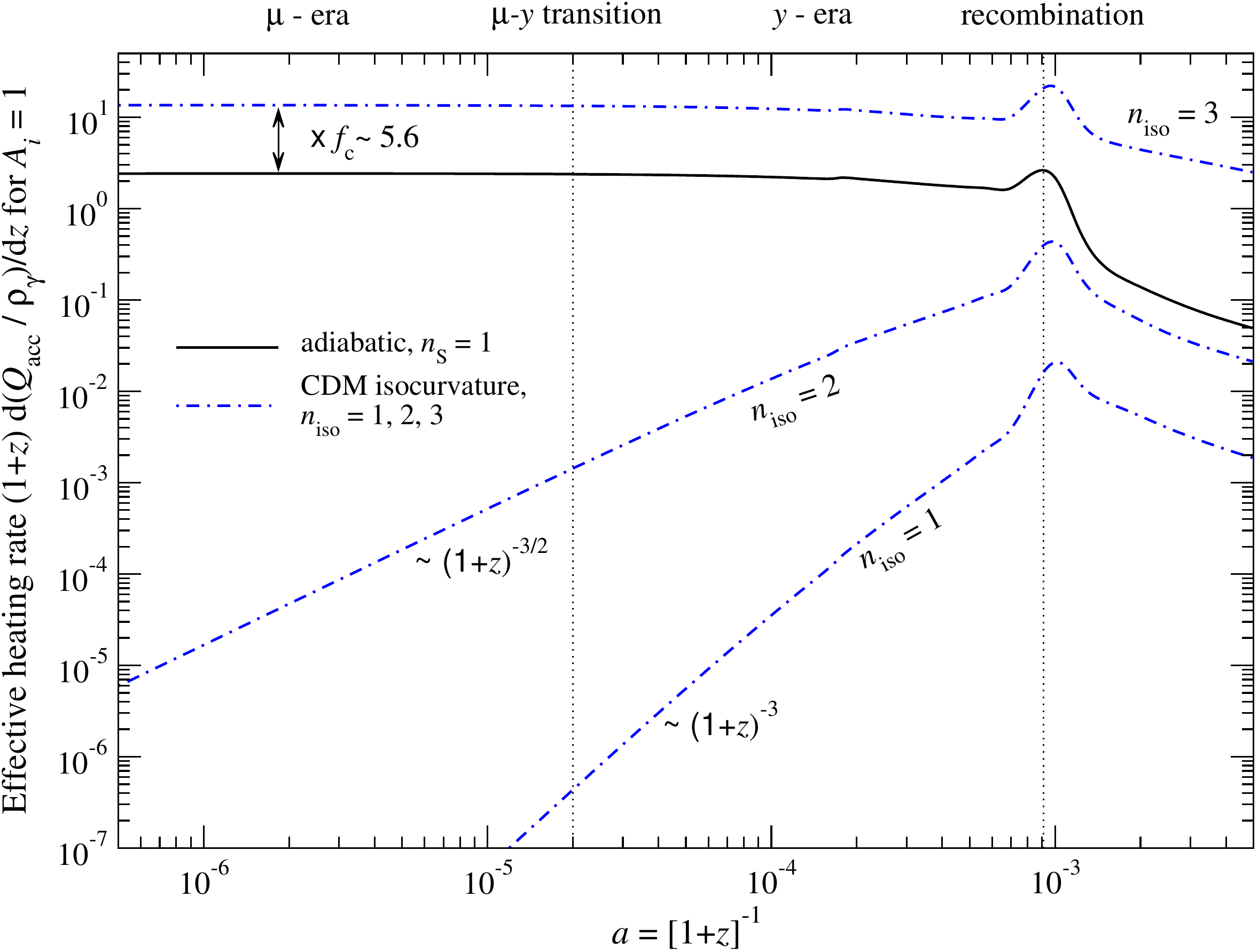}
\caption{Effective heating rate, $\id (Q_{\rm ac}/\rho_\gamma)/\id z$, for different {\it pure} perturbation modes. 
We multiplied by $(1+z)$ and set the overall amplitude of the perturbation power spectrum to unity, i.e. $A_i=1$. We also used $n_{i, \rm run}=0$ in all cases. For the upper panel, we used spectral index $n_i=1$ while in the lower we varied it for the CDM isocurvature modes as labeled. 
The annotated factors are roughly $f_{\rm cb}\simeq (\Omega_{\rm c}/\Omega_{\rm b})^2$, $f_{\rm c}\simeq (3/8)(\Omega_{\rm c}/\Omega_{\rm m})^2 (k_{\rm eq}/k_0)^2[1-(4R_\nu/15)^2]^2$, $f_{\nu}\simeq 25(1+3R_\nu/5)^{2}/[4(R_\nu/R_\gamma)^2(1+4R_\nu/15)^2]$ and $f_{\rm d}\simeq 27(1+3R_\nu/5)^{2}/[4(1+6R_\nu/5)^{2}]$. These illustrate the relative heating efficiencies for different perturbation modes with the same overall amplitude. The results were obtained by direct integration of the perturbation equations using {\sc CosmoTherm}.
}
\label{fig:heating_comp}
\end{figure*}

\section{Heating rates for pure modes}
\label{sec:heating_rates}
We now compute the effective heating rates for different perturbation modes using \textsc{CosmoTherm}, evaluating Eqs.~\eqref{eq:Sac_full} and \eqref{eq:Q_ac}, and comparing with results obtained with the analytic approximations [Eqs.~\eqref{eq:theta1_tc}, \eqref{eq:Sac_tc_new} and mode amplitudes of Sect.~\ref{sec:C_i}]. To initialize conformal Newtonian (CN) gauge fluid and metric variables in \textsc{CosmoTherm} for the different perturbation modes, we require super-horizon power-series solutions for each mode, as a function of conformal time $\eta$. Using a matrix normal-mode analysis followed by a gauge transform, we obtain these power-series solutions\footnote{To our knowledge, this is the first published summary of early-time CN gauge super-horizon solutions for this system which includes all fluid variables, in addition to the metric perturbations.} in Appendix~\ref{app:initial}. 

In Fig.~\ref{fig:heating_comp}, we show our numerical results for {\it pure} adiabatic, BI/CI and NDI/NVI  modes. We assumed a scale-invariant perturbation power spectrum with normalization $A_i=1$ to allow a comparison of the different heating efficiencies. For the CI mode, we also varied the spectral index, $n_i$. A cosmology with $\Omega_{\rm m}=0.26$, $\Omega_{\rm b}=0.044$, $\Omega_{\rm k}=0$, $\Omega_{\Lambda}=0.74$, $h=0.71$, $\Yp=0.24$, $N_{\rm eff}=3.046$, $T_0=2.726\,{\rm K}$ and improved recombination history \citep{Chluba2010b, Yacine2010c} was used in all cases.

Since for scale-invariant curvature perturbations the effective heating rate caused by adiabatic perturbations scales roughly as $\id(Q_{\rm ac}/\rho_\gamma)/\id z \propto (1+z)^{-1}$ at early times \citep[e.g., see][]{Khatri2011BE}, we multiplied all rates by $(1+z)$.
As Fig.~\ref{fig:heating_comp} shows, for scale-invariant primordial fluctuations not only for AD but also for NDI/NVI modes, this means $(1+z)\id(Q_{\rm ac}/\rho_\gamma)/\id z\simeq {\rm const}$ before recombination. 
The sub-horizon mode amplitudes for AD, NDI and NVI modes only depend on $R_\nu$. As a result, $C^2 \simeq \rm const$ [cf. Eqns.~\eqref{eq:C_ad}, \eqref{eq:C_iso_neut}, and \eqref{eq:C_iso_neut_velo}]. Using Eqs.~\eqref{eq:Q_ac} and \eqref{eq:Sac_tc_new}, the $\left(1+z\right)^{-1}$ redshift dependence of these heating rates can thus be easily derived.
On the other hand, BI and CI modes show a steeply increasing heating rate towards lower redshift.
This is because for $n_i\simeq 1$, the amplitude of the photon temperature perturbations is suppressed by $\simeq k_{\rm eq}/k$ [see Eq.~\eqref{eq:C_iso}] at early times, and $C^2(k)$ scales like $\simeq (k_{\rm eq}/k)^2$. Consequently, for $n_i=3$ (i.e. with strongly increased small-scale density perturbations) one again expects a nearly constant heating rate $(1+z)\id(Q_{\rm ac}/\rho_\gamma)/\id z \simeq {\rm const}$, as found in our calculation (cf. the lower panel of Fig.~\ref{fig:heating_comp}).

The dependence of the pre-recombination heating rate on the type of perturbation can be captured by substituting the fits of Sec.~\ref{sec:C_i} into Eq.~\eqref{eq:Sac_tc_new} and defining an effective spectral index $n_i^\ast$, yielding
\bsub
\label{eq:coeffies_Qdot}
\beal
D^2& \simeq\frac{1}{(1+4R_\nu/15)^2}
\simeq 0.81 &n^\ast_i&=n_i & \text{(AD)}
\\
D^2& \simeq\frac{3}{8}
\left(\frac{\Omega_{\rm b}\,k_{\rm eq}}{\Omega_{\rm m}\,k_0}\right)^2 \beta_\nu
\simeq 0.19
&n^\ast_i&=n_i-2 &\text{(BI)}
\\
D^2& \simeq\frac{3}{8}
\left(\frac{\Omega_{\rm c}\,k_{\rm eq}}{\Omega_{\rm m}\,k_0}\right)^2 \beta_\nu
\simeq 4.7
&n^\ast_i&=n_i-2 &\text{(CI)}
\\
D^2& \simeq\frac{4(R_\nu/ R_\gamma)^2}{25(1+3R_\nu/5)^{2}} 
\simeq 0.05
&n^\ast_i&=n_i & \text{(NDI)}
\\
D^2& \simeq\frac{27(R_\nu/R_\gamma)^2}{25(1+6R_\nu/5)^{2}} 
\simeq 0.23
&n^\ast_i&=n_i & \text{(NVI)}
\end{align}
\esub
with $\beta_\nu=(1-4R_\nu/15)^{2}\simeq 0.79$. At early times, we therefore have the effective heating rates
\beal
\label{eq:Q_ac_eff}
\frac{1}{a^4 \rho_\gamma}\frac{\id (a^4 Q_{\rm ac})}{\id z}
& \approx 
2 D^2  \int \mathcal{P}^\ast_i(k) \, \partial_z \expf{-2k^2/\kD^2} \id\ln k
\end{align}
where $\mathcal{P}^\ast_i(k)\equiv A_i \,(k/k_0)^{n^\ast_i-1+\frac{1}{2} n_{i, \rm run} \ln(k/k_0)}$. 
We confirmed numerically that at redshifts $z\gtrsim 10^4$ these approximations work pretty well, giving $\simeq 10\%-15\%$ precision for the effective heating rate.
Here, $\varepsilon = 2D^2\simeq \rm const$ defines a mode dependent heating efficiency.
This implies that the early SDs produced by the different modes considered here are all degenerate with an overall normalization when comparing AD, NDI and NVI for $n^\ast_i=n_i$ on one hand, with BI and CI modes for $n^\ast_i=n_i-2$ on the other.
The differences derive from how much of the initial perturbations in the different fluid variables at small scales actually appear as perturbations in the photon field.

Comparing the heating efficiencies, Eq.~\eqref{eq:coeffies_Qdot}, shows that AD modes dissipate their energy roughly 16 times more efficiently than NDI fluctuations. Similarly, NVI modes have $\simeq 4.7$ times higher heating efficiency than NDI modes.
Furthermore, BI modes source early SDs at about $(\Omega_{\rm c}/\Omega_{\rm b})^2\simeq 24$ lower efficiency than CI modes, while in comparisons to AD modes CI fluctuations for $n^\ast_i=n_i-2$ cause $\simeq 5.6$ times larger heating.
All these statements are confirmed by our numerical results (cf. Fig.~\ref{fig:heating_comp}).

Closer to the recombination epoch baryon loading no longer is negligible and we see a suppression of the heating rate relative to the high-redshift scaling (cf. Fig.~\ref{fig:heating_comp}). 
After the recombination epoch ($z\lesssim 1000$), the effective heating rates drop significantly as photons begin free streaming. At this late stage, the second-order Doppler effect starts contributing significantly \citep{Chluba2012}. 
For the baryon and CDM isocurvature modes, the post-recombination heating rate is relatively larger than for the adiabatic case, emphasizing the importance of late bulk motions for these perturbation modes. 
The neutrino isocurvature modes also result in relatively less heating at the late stages.
This implies that the relative ratio of $\mu$- and $y$- parameters is slightly mode dependent; however, from the observational point of view, the differences are too small and degenerate with the shape of the power spectrum itself to allow distinguishing different scenarios in a model-independent way.

\subsection{Approximations for power-law perturbation spectra}
One can gain additional insight by considering primordial fluctuations with power-law perturbation power spectra ($n_{\rm run, i}=0$). From Eq.~\eqref{eq:Q_ac_eff}, it is straightforward to show that the pre-recombination heating rate is roughly given by \citep[cf.][for adiabatic modes]{Chluba2012}:
\beal
\label{eq:heat_SZ_appr_final_nS}
\frac{1}{a^4 \rho_\gamma}\frac{\id (a^4 Q_{\rm ac})}{\id z}
\approx 
\frac{3 A_i\, D^2 }{1+z} 
\;\Gamma\left(\frac{1+n^\ast_i}{2}\right)
\left[ \frac{(1+z)^3}{2k_0^2 A_{\rm D}}\right]^{\frac{n^\ast_i-1}{2}}.
\end{align}
where $D^2$ and $n^\ast_i>-1$ are mode dependent [see Eq.~\eqref{eq:coeffies_Qdot}] and 
\beal
\nonumber
A_{\rm D}\approx \frac{ (16/15) \, c } {18 H_0 \Omega^{1/2}_{\rm r} N_{\rm e, 0} \sigT} \approx 
\pot{5.92}{10}\,{\rm Mpc^{2}}.
\end{align}
Here, $H_0$ is the Hubble parameter, $\Omega_{\rm r}$ is the density of relativistic species and $N_{\rm e}=N_{\rm e, 0}(1+z)^3$ is the number density of electrons (bound and free).
Equation~\eqref{eq:heat_SZ_appr_final_nS} explicitly shows that for AD, BI, CI, NDI, and NVI modes and $n^\ast_i\equiv 1$ the heating rate indeed scales like $\id(Q_{\rm ac}/\rho_\gamma)/\id z \propto (1+z)^{-1}$.
In terms of the power spectrum, $P_i(k)$, this mean scale-invariant perturbations for the AD, NDI, and NVI modes, while for BI and CI mode a very blue primordial power spectrum with $n_i=3$ is necessary.
Also, for BI and CI fluctuations with spectral index $n_i\simeq 1$ ($n^\ast_i \simeq -1$), from Eq.~\eqref{eq:heat_SZ_appr_final_nS} we find a redshift scaling $\id(Q_{\rm ac}/\rho_\gamma)/\id z \propto (1+z)^{-4}$, while for $n_i=2$ ($n^\ast_i=0$) one has $\id(Q_{\rm ac}/\rho_\gamma)/\id z \propto (1+z)^{-5/2}$.
Our numerical calculations confirm this dependence (cf. Fig.~\ref{fig:heating_comp}); these scalings are, however, only valid at early time, before the recombination era ($z\gtrsim 10^4$).
Especially, when $n_i\simeq 1$, large-scale modes contribute strongly to the total heating integral. The transfer functions of these modes are not well represented by the simple approximations given above, so that the integral formally diverges, unless a cutoff is introduced at small $k$.
In this case, one has to resort to full numerical integration of the perturbation equations. 
Still for $n_i>1$, Eq.~\eqref{eq:heat_SZ_appr_final_nS} provides a fairly accurate estimate for the effective heating rate.

Equation~\eqref{eq:heat_SZ_appr_final_nS} also shows that the main dependence of the heating rate (for $n_{i, \rm run}=0$) for the adiabatic and neutrino isocurvature modes is due to the overall amplitude of the perturbations power spectrum, $A_i$, the spectral index, $n_i$, and the value of $R_\nu$, which changes the heating efficiency, $\varepsilon\simeq 2D^2$. The dependence on the dissipation scale (related to $A_{\rm D}^{-1/2}\simeq \pot{4.1}{-6}\,\Mpc^{-1}$) is much weaker unless the small-scale spectral index, $n_i$, differs strongly from unity.
For the BI and CI modes $\Omega_{\rm b}$ and $\Omega_{\rm c}$ as well as $k_{\rm eq}$ become important. Cosmology dependence introduced by the dissipation scale is again less important for $n_i\simeq 3$.

\section{Spectral distortion constraints on early-universe cosmology}
\label{modelsec}
At large scales, $k\lesssim 1\,\Mpc^{-1}$, constraints derived from CMB anisotropies and large-scale structure measurements are pretty tight already, suggesting nearly scale-invariant adiabatic perturbation with amplitude of the primordial curvature power spectrum $A_\zeta\simeq \pot{2.4}{-9}$ \citep{Dunkley2010, spt, Planck2013params}. On the other hand, at small scales constraints are much weaker \citep[see][for some discussion]{BSA11} and SDs provide a complementary (if not the only), strong probe for the primordial power spectrum.

The discussion of Sect.~\ref{sec:heating_rates} already indicates that the heating caused by very different perturbations modes and their mixtures can lead to similar distortions. 
For example, given $P_i(k)$ the time dependence of the heating rates for AD, NDI and NVI modes implies that from the practical point of view the SD signal should be indistinguishable, up to an overall efficiency factor that is degenerate with the power spectrum amplitude.
Similarly, the SD arising from dissipation of BI and CI modes should be practically indistinguishable (small differences might arise in the post-recombination epoch, where non-linear effects will become important).
Comparing the SD from AD, NDI and NVI modes with those from BI and CI modes for given primordial power spectrum the former produce a much larger $\mu$ distortion due to the $\simeq k_{\rm eq}/k$ suppression of the BI/CI mode amplitude. This is, however, degenerate with the overall spectral index of the perturbations.
Still, on a model-by-model basis interesting constraints on the small-scale power spectrum can be derived for each case, as we illustrate here.

\subsection{Estimates for the $\mu$- and $y$-parameters and the definition of $k$-space window functions for pure modes}
\label{sec:estimates}
We already showed that the pre-recombination heating rates from different perturbations can all be represented by one single expression, Eq.~\eqref{eq:Q_ac_eff}, with heating efficiency, $\varepsilon\simeq 2 D^2$, according to Eq.~\eqref{eq:coeffies_Qdot}. For power-law power spectrum one obtains the compact expression, Eq.~\eqref{eq:heat_SZ_appr_final_nS}.
Here, we generalize to arbitrary shapes of the primordial power spectrum and estimate the SD by computing the redshift integrals, Eq.~\eqref{eq:def_mu_y_Dr_r}.
For the $y$-parameter the integral can be performed analytically, while for the $\mu$-parameter the spectral distortion visibility function, $\mathcal{J}_{\rm bb}(z)$, requires numerical integration.
This reduces the problem to a 1-dimensional integral over $k$-space window functions for the effective $\mu$- and $y$-parameters:
\bsub
\label{eq:mu_y}
\beal
\mu_{\rm ac}&\approx  \int_{k_{\rm min}}^\infty \frac{k^2 \id k}{2\pi^2} P_i(k) \, W^{\mu}_i(k)
\\[1mm]
y_{\rm ac}&\approx  \int^\infty_{k_{\rm min}} \frac{k^2 \id k}{2\pi^2} P_i(k) \, W^{y}_i(k),
\end{align}
\esub
where the $k$-space window functions are
\bsub
\label{eq:window_def}
\beal
W^{\mu}_i(k)&\approx  
2.8 \,C^2(k) \int_{\zmuy}^\infty \mathcal{J}_{\rm bb}(z)\,\partial_z \expf{-2k^2/\kD^2} \id z
\\[1mm]
\nonumber
&\approx  2.8 \,C^2(k) \left[
\exp\left(-\frac{\left[\frac{\hat{k}}{1360}\right]^2}{1+\left[\frac{\hat{k}}{260}\right]^{0.3}+\frac{\hat{k}}{340}}\right) 
- \exp\left(-\left[\frac{\hat{k}}{32}\right]^2\right)
\right]
\\[2mm]
W^{y}_i(k)&\approx  \frac{C^2(k)}{2}\,  
\expf{-2k^2/\kD^2(\zmuy)}
\approx  \frac{C^2(k)}{2}\,  \exp\left(-\left[\frac{\hat{k}}{32}\right]^2\right),
\end{align}
\esub
with $\hat{k}=k/[1\,\Mpc^{-1}]$ and cutoff scale, $k_{\rm min}\simeq 1\,\Mpc^{-1}$, both because modes at $k<1\, \Mpc^{-1}$ are already tightly constrained by CMB measurements at large scales, and because the analytic approximations for the photon transfer functions introduced above become inaccurate.
The approximations for $W(k)$ are for the concordance cosmology. 
These expressions are similar to those given by \citet{Chluba2012inflaton}, but for $W^{\mu}_i(k)$ we matched the numerical results well not only in the small and large $k$ limits but also at intermediate scales.
In Eq.~\eqref{eq:window_def}, the $k$-dependent factor $C^2(k)$ depends on the perturbation type. 
For AD and NDI/NVI modes one has $C^2(k)=D^2$, while for BI/CI modes $C^2(k)=D^2(k)\,(k_0/k)^2$ with the values of $D^2$ given by Eq.~\eqref{eq:coeffies_Qdot}.
The values of $\mu$- and $y$-parameters for different modes can thus be estimated knowing $D^2$ and rescaling the primordial power spectrum by appropriate powers of $k_0/k$. Mode mixtures can be treated in a similar way (see Sect.~\ref{sec:mode-mix}).

\begin{figure*}
\centering
\includegraphics[width=1.28\columnwidth]{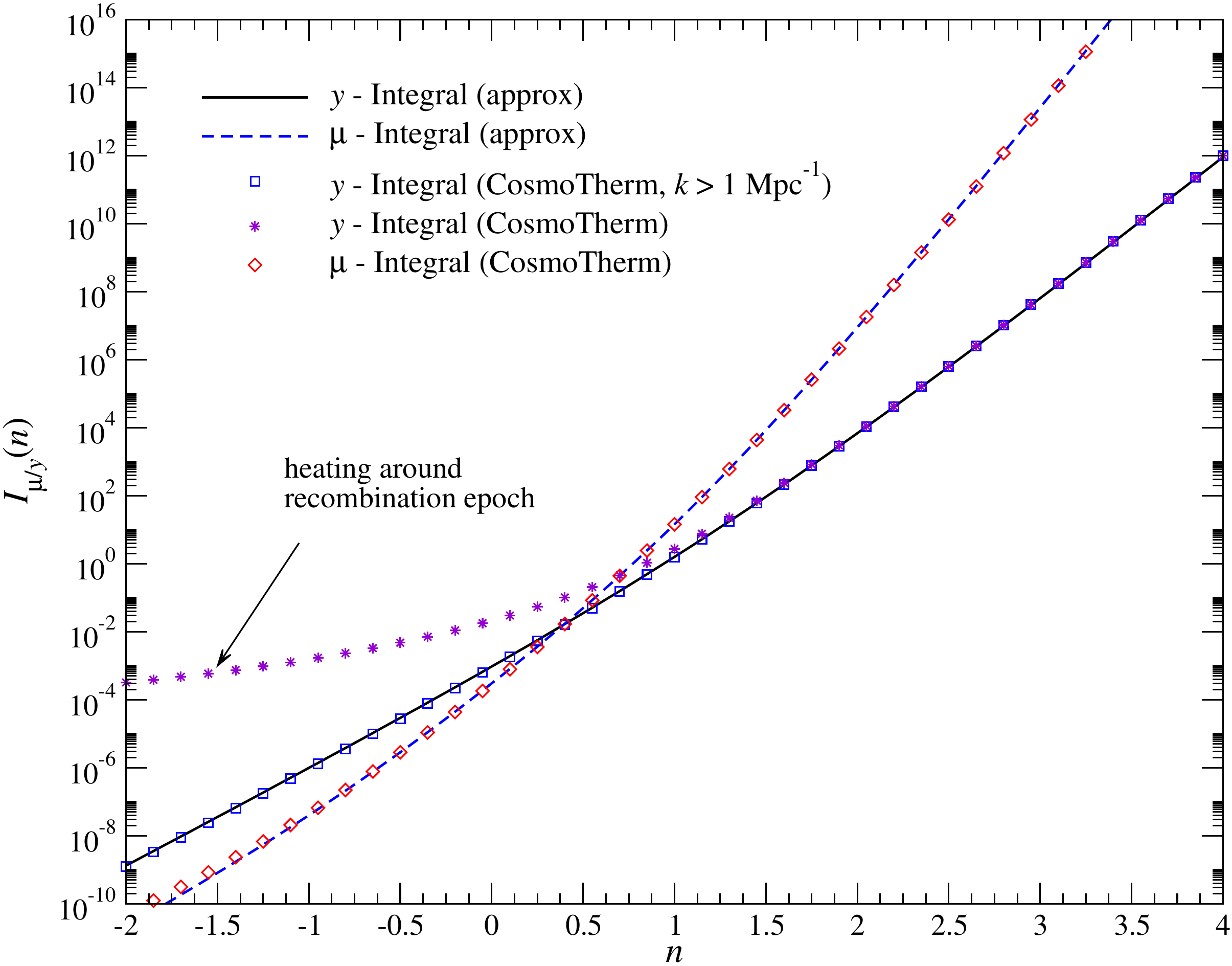}
\caption{Dependence of the heating integrals $I_\mu(n)$ and $I_y(n)$ on the spectral index, $n$. We assumed pivot scale, $k_0=0.002\,\Mpc^{-1}$, to make the spectral distortion constraint directly comparable with the large-scale CMB constraint; values for $k^\ast_0\neq k_0$ can be obtained by rescaling with $(k^\ast_0/k_0)^{1-n}$. For comparison, we also give the results for the heating integrals obtained with {\sc CosmoTherm} for the adiabatic modes. We confirmed that the integrals for all the different modes considered here agree with each other to high precision.}
\label{fig:heating_Ints}
\end{figure*}

\subsection{Constraints on different pure perturbation modes}
\label{sec:diss_constr}
In this section, we highlight constraints derived from SD measurements of {\it COBE}/FIRAS and future {\it PIXIE}-type experiments, focusing the discussion on pure perturbation modes. The derived limits should be interpreted as {\it conservative} upper bounds, since not only can several types of perturbation modes be present at small scales, but also other sources of early energy release (e.g., decaying or annihilating relics, superconducting cosmic strings) could increase the CMB distortion. This would generally tighten the constraint on each source of energy injection.

\subsubsection{Pure AD, NDI and NVI modes}
For AD modes, a detailed discussion of SD power spectrum constraints derived from {\it COBE}/FIRAS and a {\it PIXIE}-type experiment can be found in \citet{Chluba2012inflaton}. 
Since the pre-recombination heating rate for NDI and NVI modes only differs by an overall efficiency factor from the one of AD modes, their analysis directly carries over.
For the NDI mode, the relative heating efficiency is $f_{\nu}\simeq 25(1+3R_\nu/5)^{2}/[4(R_\nu/R_\gamma)^2(1+4R_\nu/15)^2]\simeq 16$, which can be captured by replacing $A_\zeta \rightarrow A_\zeta/f_\nu \simeq 16\, A_\zeta$ in the work of \citet{Chluba2012inflaton}.
Similarly, for the NVI mode, we have $f \simeq 27(R_\nu/R_\gamma)^2(1+4R_\nu/15)^2/[25(1+6R_\nu/5)^{2}]\simeq 0.29$ and $A_\zeta \rightarrow  3.5\, A_\zeta$. Both for NDI and NVI modes, the constraints are thus weaker than for the AD mode.

To give some examples, for AD modes {\it PIXIE} is able to rule out a scale-invariant curvature power spectrum with $A_{\zeta} \gtrsim \pot{4.3}{-9}$ at wavenumber $k>50\,\Mpc^{-1}$ with $5\sigma$ confidence if no $\mu$-type distortion is detected \citep{Chluba2012, Chluba2012inflaton}.
This therefore means that {\it PIXIE} would also be able to rule out scale-invariant neutrino density perturbations with overall amplitude $A_i\gtrsim\pot{7.0}{-8}$ at wavenumber $k>50\,\Mpc^{-1}$ with $5\sigma$ confidence.
Even the $\mu$-limit from {\it COBE}/FIRAS already implies $A_i\lesssim\pot{1.3}{-4}$ at $2\sigma$ level.
For the NVI mode, these amplitude constraints are $\simeq 4.7$ times tighter.

\subsubsection{Pure BI and CI modes}
Also for BI and CI modes, the analysis of \citet{Chluba2012inflaton} can be directly applied; however, not only are overall factors of $f_{\rm b}\simeq 0.23$ and $f_{\rm c}\simeq 5.6$ needed, respectively, for the conversion of different constraints on adiabatic modes, but also one must use the effective spectral index $n^\ast_i=n_i-2$.
For instance, for $n^\ast_i=1$ {\it PIXIE} would be sensitive to $A_{\rm b}\gtrsim \pot{1.9}{-8}$ for the baryon, and $A_{\rm c} \gtrsim\pot{7.3}{-10}$ for the CI modes at wavenumber $k>50\,\Mpc^{-1}$ with $5\sigma$ confidence.
The limits from {\it COBE}/FIRAS are about 1800 times weaker at $2\sigma$ level.
This value of $n^\ast_i$ means a very blue small-scale perturbation spectrum with spectral index $n_i=3$. This blue spectrum can be realized in certain axion isocurvature models, for example, in which the Peccei-Quinn symmetry breaking scale is itself dynamical during inflation \citep{Kasuya2009}, as noted in \citet{Dent2012}. For $n_i=1$ the constraints are much weaker, as shown below.

\subsubsection{Simple expressions for power-law perturbation spectra}
For pure power-law primordial spectra, $\mathcal{P}_i=A_i \,(k/k_0)^{n_i-1}$, we can simplify the computation of limits from $\mu$ and $y$ significantly using the $k$-space window functions, Eq.~\eqref{eq:mu_y}.
Defining the heating integrals 
\bsub
\label{eq:J_mu_y}
\beal
I_\mu(n)&=  \int_{k_{\rm min}}^\infty (k/k_0)^{n-1} \, \hat{W}^{\mu}_i(k) \id\ln k
\\
I_y(n)&=  \int^\infty_{k_{\rm min}}  (k/k_0)^{n-1}\, \hat{W}^{y}_i(k) \id\ln k ,
\end{align}
\esub
with $\hat{W}^{\mu/y}_i(k)=W^{\mu/y}_i(k)/D^2$ and $k_{\rm min}=1\,\Mpc^{-1}$, limits on the overall amplitude of the power spectrum at small scales ($k\gtrsim 1\,\Mpc^{-1}$) derived from $\mu$ and $y$ distortions can be expressed as
\beal
\label{eq:mu_y_constraint}
A_i &\lesssim  \frac{\mu_{\rm lim}}{D^2 I_\mu(n^\ast_i)}, 
&
A_i &\lesssim  \frac{y_{\rm lim}}{D^2 I_y(n^\ast_i)}.
\end{align}
The $y$-limit probes power at scales $1\,\Mpc^{-1} \lesssim k \lesssim 50\,\Mpc^{-1}$, while the $\mu$-limit is most sensitive to scales $50\,\Mpc^{-1}\lesssim k \lesssim 10^4\,\Mpc^{-1}$. Late energy release ($z\lesssim 10^4$), during and past the recombination epoch, is not included here. The corresponding $y$-distortion can, however, be computed using {\sc CosmoTherm} in that case.

In Fig.~\ref{fig:heating_Ints}, we illustrate the dependence of $I_\mu(n^\ast_i)$ and $I_y(n^\ast_i)$ on the spectral index.
These functions have a very steep dependence on $n^\ast_i$, which in the range $-2<n^\ast_i<5$ can be approximated by
\beal
\label{eq:J_mu_y_approx}
\ln I_\mu(n)&\approx 
2.73\left[ 
1 + 4.42 \xi + 0.444 \xi^2 - \pot{9.21}{-3} \xi^3 
\right.
\nonumber\\
&\qquad \qquad\left. - 0.0168 \xi^4 - \pot{5.38}{-5} \xi^5 + \pot{4.92}{-4} \xi^6\right]
\nonumber\\[1mm]
\ln I_y(n)&\approx 
0.504\left[ 1 + 15.66 \xi + 0.845 \xi ^2 + 0.0253 \xi ^3 - 0.0189 \xi ^4\right],
\nonumber
\end{align}
with $\xi =n-1$. These expressions provide a $5\%-10\%$ fit to the numerical results for Eq.~\eqref{eq:J_mu_y}. For $n=1$, we find $I_\mu(1)\approx 14.4$ and $I_y(1)\approx 1.59$ numerically.

We can directly check the precision of these approximations for the heating integrals using {\sc CosmoTherm}. The results are also shown in Fig.~\ref{fig:heating_Ints}.
For the $\mu$-integral, the approximations work very well. The approximations for the $y$-integral represent the full numerical result well if dissipation of modes at $k<1\,\Mpc^{-1}$ is neglected (consistent with the approximations made above). 
At larger scales, the approximations for the transfer functions are not valid, since the tight-coupling limit breaks down. Also, baryon loading and the second-order Doppler effect become important so that we can only compute the effect accurately using {\sc CosmoTherm}. 
Our result shows that a significant amount of energy is dissipated at late times if $n_i\lesssim 1$.
These results are especially important for the baryon and CDM isocurvature modes, which for scale-invariant primordial density power spectrum would have $n^\ast_i=-1$. This renders a constraint derived from the $y$-parameter about $\simeq 1500$ times tighter than the limit obtained by only accounting for modes with $k>1\,\Mpc^{-1}$ and assuming that the power-law spectrum goes all the way from CMB scales to small-scales.
These limits are still not competitive with those derived at CMB scales. For instance, from {\it COBE}/FIRAS we have $|y|<\pot{1.5}{-5}$. For $n^\ast_i=-1$ this means $A_{\rm b}\lesssim \pot{5}{-2}$ for the baryon and $A_{\rm c} \lesssim\pot{2}{-3}$ at $2\sigma$ confidence. With {\it PIXIE} this could be improved by a factor of $\simeq 1800$, but this is still far away from current constraints.

For power spectra with $n^\ast_i > 1$, constraints from $\mu$ are generally expected to be tighter than those obtained with $y$.
Also, since there are many sources of large $y$-distortions at low redshift, 
a clean detection of the primordial signal will be very challenging.
In all these cases one should consider the possibilities to include spatial-spectral and polarization information to separate the different components. 
For example, the patchiness of the $y$-distortion \citep[see][for the first all sky $y$-map]{Planck2013ymap} from reionization depends on the characteristic bubble size \citep{Zhang2004}, which might allow primordial and reionization-induced $y$-distortions to be distinguished.
A detailed discussion of this possibility is beyond the scope of this paper.

\begin{figure*}
\centering
\includegraphics[width=1.2\columnwidth]{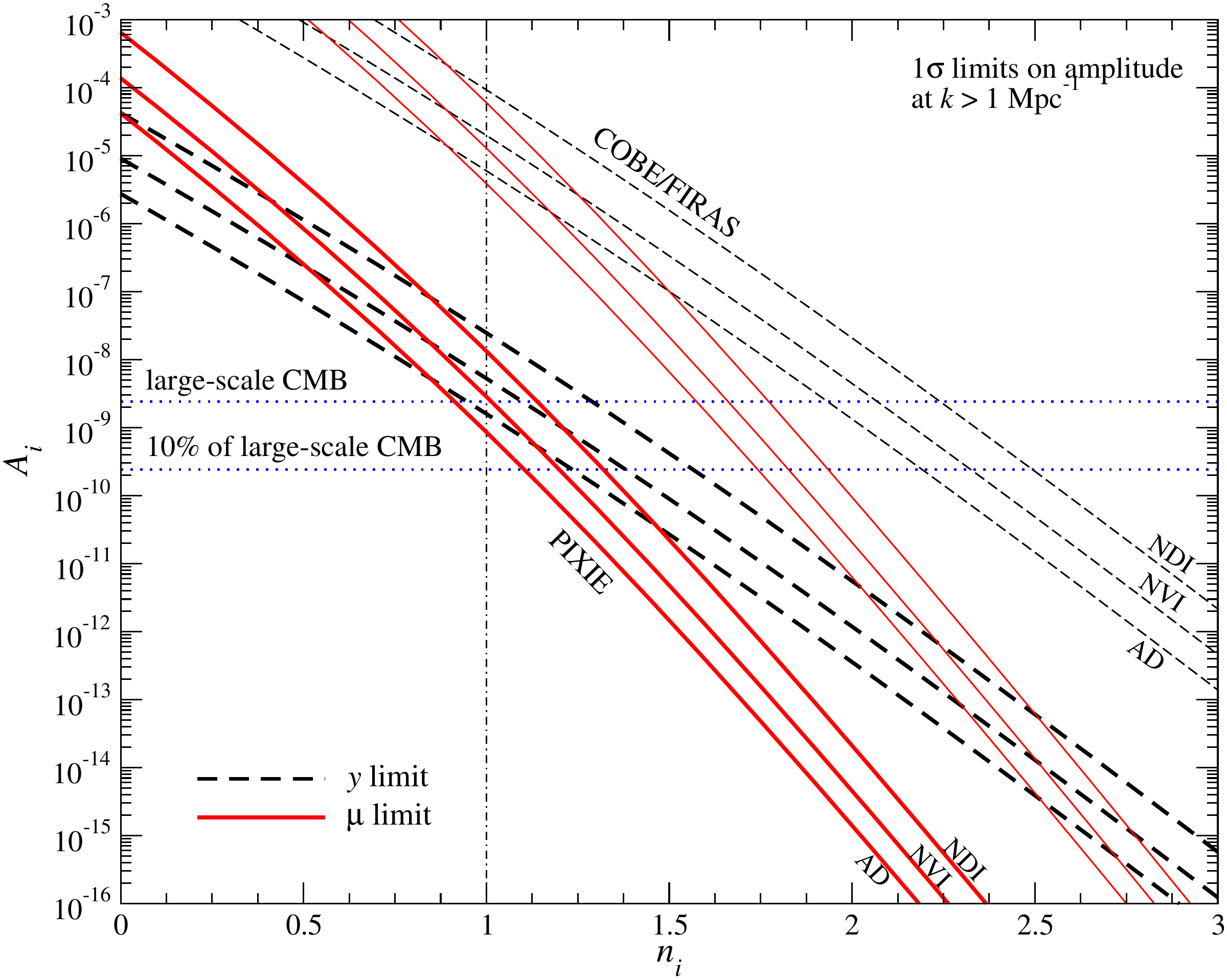}
\\[1mm]
\includegraphics[width=1.2\columnwidth]{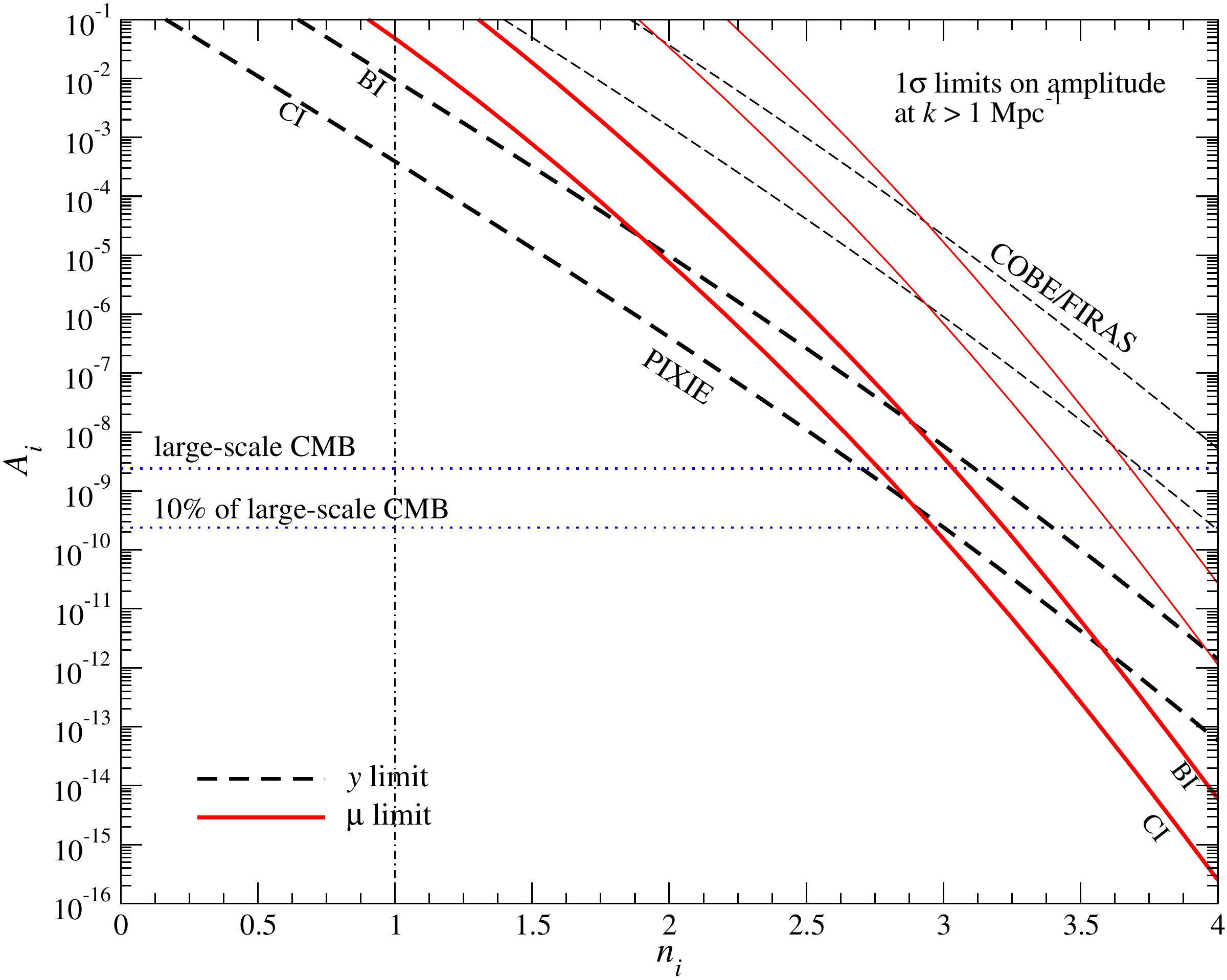}
\caption{Limits on the amplitude of the power spectrum at $k>1\,\Mpc^{-1}$ for different pure perturbations modes and spectral indices. The heavy lines show constraints for a {\it PIXIE}-type experiment with $1\sigma$ detection limits $y=\pot{2}{-9}$ and $\mu=10^{-8}$. Light lines are present limits from {\it COBE}/FIRAS. Mode amplitudes above the corresponding lines are/will be ruled out by CMB spectral distortion measurements. 
Assuming one overall power-law perturbation spectrum at small scales, the limits derived from $\mu$ and $y$ are not independent, and their ratio can in principle be used to distinguish AD, NDI and NVI on the one side from BI and CI, on the other (see the text for discussion). 
Interpreting the limits independently, $y$-distortions constrain power at $1\,\Mpc^{-1}\lesssim k\lesssim 50\,\Mpc^{-1}$, while the limit from $\mu$ probes power at  $50\,\Mpc^{-1}\lesssim k\lesssim 11000\,\Mpc^{-1}$.
Note also that we assumed pivot scale, $k_0=0.002\,\Mpc^{-1}$, to make the spectral distortion constraint directly comparable with the large-scale CMB constraint; values for $k^\ast_0\neq k_0$ can be obtained by rescaling with $(k^\ast_0/k_0)^{1-n}$.}
\label{fig:limits}
\end{figure*}

Finally, in Fig.~\ref{fig:limits} we illustrate $1\sigma$ constraints derived from {\it COBE}/FIRAS and {\it PIXIE} in the $A_i-n_i$ plane. These were computed using Eqs.~\eqref{eq:mu_y_constraint} and \eqref{eq:J_mu_y_approx}.
Putting experimental obstacles aside, for AD, NDI and NVI at $n_i>0.9$ the limits from $\mu$-distortions are tighter than those derived from a measurement of the $y$-parameter. For BI and CI modes, this transition occurs at $n_i\simeq 2.9$. Furthermore, due to the steep dependence on $n_i$, constraints on the small-scale amplitude of AD, NDI and NVI modes become very tight at $n_i>1$, showing the impressive potential of testing early-universe models that produce large excess small-scale power.

\subsection{Constraints on mixed modes}
\label{sec:mode-mix}
For the discussion in the previous section, we assumed that only one type of perturbation was present. Constraints from CMB distortions on the amplitude of different modes of course can only limit the total energy release by the mode mixture. The heating efficiencies can be obtained as explained in Sect.~\ref{sec:A_mixed}. If correlations between modes are neglected, the final heating rates can be described using Eq.~(\ref{eq:Q_ac_eff}) and adding up the contributions of different modes.

Mode mixtures can in principle give rise to interesting behavior. For example, a CI/BI isocurvature mode with a very blue spectral index could be mixed with a quasi-scale invariant AD mode. This means that at small scales the isocurvature mode could dominate, while at large scales, which are well constrained by CMB experiments, the AD mode is most important.

A similar story could be told for the NDI or NVI modes mixed with AD modes, and the constraints for these cases can be directly deduced from Fig.~\ref{fig:limits}. If on the other hand the isocurvature mode is completely sub-dominant, the SD limit will be very weak and mainly constrain the AD mode amplitude at small scales. For mixtures with comparable contributions of AD and another mode, interference terms (destructive and constructive) make things more interesting. 
For example, in the curvaton scenario, isocurvature modes are all correlated with the adiabatic mode, and the precise correlation coefficients are determined by the curvaton energy density fraction $r_{\rm D}$ at curvaton decay, the lepton asymmetry parameter (which sets the number of effective relativistic degrees of freedom $N_{\rm eff}$), and whether dark matter density, lepton number and baryon number are produced before, by, or after curvaton decay.

To determine whether or not curvaton-generated SDs could ever be detected, we sweep through the model space of $27$ ($3^3$) permutations, allowing the neutrino asymmetry parameter to span the entire (rather permissive) allowed experimental range. We generalize Eqs.~(\ref{eq:C_mix_correlated}) to allow all possible correlated mixtures of adiabatic and isocurvature modes, and use the WKB coefficients of Sec. \ref{sec:C_i}. We self-consistently compute these in the presence of a lepton asymmetry, applying expressions in \citet{2002PhLB..524....5L} and \citet{2003PhRvD..67b3503L}. To properly compute the curvaton-generated $y$ distortion, we use the numerically determined $I_y(n_{i}^{*})$.

The curvaton model would also seed local-type non-Gaussianity with amplitude $f_{\rm NL}^{\rm local}$. We thus restrict consideration to the range $r_{\rm D}\geq 0.15$, the range still allowed by \textit{Planck} constraints to $f_{\rm NL}^{\rm local}$ \citep{Planck2013ng}. Additional limits to $r_{\rm D}$ could be obtained more directly from \textit{Planck} constraints to isocurvature modes, but these constraints have not been established in the presence of all four isocurvature modes with general correlations \citep{Planck2013iso}. Even though the curvaton model can excite the NDI mode, the overall change in $\mu$ and $y$ over the null (adiabatic) hypothesis is of the order of $10\%$, and thus undetectable at the sensitivity level possible with {\it PIXIE}. Future advances may change this dim state of affairs. An in-depth discussion of other correlated models is beyond the scope of this paper.

\section{Conclusions}
\label{sec:conclusions}
In the future, spectral distortions of the CMB might provide a powerful new probe of early-universe physics. Here, we studied distortions produced by the dissipation of small-scale perturbations, exploring the dependence of the signal on the different types of cosmological initial conditions.
As one main result, we obtained a unified formalism for the specific heating rates of the modes, allowing us to describe the effect of pure modes but also mode mixtures in a quasi-analytic manner (see Sects.~\ref{sec:C_i} and \ref{sec:heating_rates}).
Our expressions can be used for precise computations of the SD signal using {\sc CosmoTherm} to make more detailed forecasts, although here we restrict our attention to estimates of the associated $\mu$- and $y$-parameters, providing a simple way to constrain different early-universe models.

We find that for scale-invariant initial conditions of comparable perturbation amplitude, the heating rates from pure BI, CI and CIP fluctuations are extremely sub-dominant to those from AD modes. In agreement with \citet{Dent2012}, we show that the SD signal of CI and BI modes falls below {\it PIXIE}'s sensitivity, unless primordial perturbations have a very blue spectral index $n_{i}\gsim 3$. The BI SD signature is suppressed by $\simeq \left(\Omega_{\rm b}/\Omega_{\rm c}\right)^{2}$ compared to the CI signal, a factor that is degenerate with the overall amplitude of the power spectrum. 
The CIP SD signal is strongly suppressed in addition by cancellation, and thus is unlikely to ever be detected. This all can be understood from the fact that neither baryons nor CDM can dramatically drive the evolution of the other fluid components until after matter-radiation equality. 

The NDI and NVI modes, on the other hand, yield heating rates comparable to the AD mode at all times, as neutrinos are relativistic and similarly important to photons in driving the dynamics of the full coupled fluid system during radiation domination. 
We determined mode dependent heating efficiencies [see Eq.~\eqref{eq:coeffies_Qdot}] which weakly depend on the effective number of neutrino species\footnote{Changing effective neutrino number from $N_{\rm eff}=3$ to $N_{\rm eff}=4$ changes the heating efficiency of the adiabatic mode by $\simeq -3\%$.}, because additional non-interacting relativistic degrees of freedom carry away part of the initial perturbation power, never sourcing any perturbations in the photon fluid. Thus, SDs are in principle sensitive to the presence of dark radiation or sterile neutrinos; however, this only causes an overall normalization factor that is degenerate with the power spectrum amplitude. SDs could in principle be used to probe parameters of the curvaton scenario for primordial fluctuations, although this would require rather futuristic $10\%$ level precision in measurements of $\mu$ and $y$, which themselves are expected at the level $\Delta I/I \simeq 10^{-9}-10^{-8}$ and thus are challenging to detect.

While rather stringent limits can be derived for specific models of the small-scale power spectrum (Fig.~\ref{fig:limits}), SDs cannot tell the signature of different perturbation modes apart. For example, AD, NDI and NVI modes should all cause a very similar SD signal up to an overall efficiency factor, assuming that the power spectrum has the same shape. This is because the heating rate at different redshifts scales in practically the same way (Fig.~\ref{fig:heating_comp}).
Still, limits derived from SDs can be used as powerful tool to rule out different early-universe models, and more detailed forecasts will be necessary to demonstrate the full potential of this new window.

\small 

\section*{Acknowledgements}
JC thanks Rishi Khatri, Enrico Pajer, Richard Shaw and Rashid Sunyaev for useful discussions and comments on the paper. DG thanks Mustafa Amin, Tristan Smith and David Spergel for stimulating discussions about spectral distortions, isocurvature fluctuations and cosmic initial conditions.
The authors also acknowledge the use of the GPC supercomputer at the SciNet HPC Consortium. SciNet is funded by: the Canada Foundation for Innovation under the auspices of Compute Canada; the Government of Ontario; Ontario Research Fund - Research Excellence; and the University of Toronto. JC acknowledges support from the grants DoE SC-0008108 and NASA NNX12AE86G. DG was supported at the Institute for Advanced Study by the National Science Foundation (AST-0807044) and NASA (NNX11AF29G).

\begin{appendix}

\section{Initial conditions in the conformal Newtonian gauge}
\label{app:initial}
{\sc CosmoTherm} uses the conformal Newtonian (CN) gauge to follow the evolution of metric and fluid perturbations.
Using the conventions of \citet{Ma1995}, \citet{Bucher2000}, and \citet{Shaw2010magn}, we can determine the required initial conditions for the different perturbation modes in this gauge, working deep in the radiation dominated epoch, and in the super-horizon regime.
For convenience, we define $R_{\rm c}=\Omega_{\rm c}/\Omega_{\rm m}$, $R_{\rm b}=\Omega_{\rm b}/\Omega_{\rm m}$, $R_\gamma=1-R_\nu$, $a_\nu=1+4R_\nu/15$, $b_\nu=1+2R_\nu/5$, $c_\nu=1-4R_\nu/15$, $\omega=\Omega_{\rm m}H_0/[4\sqrt{\Omega_{\rm r}}\,c]$, and $\tau=\omega\eta$. 
According to these conventions, the scale factor evolves as $\omega \,a(\tau)=\tau+\tau^{2}$  at times when only matter and radiation are energetically relevant. In these units, the scale factor at matter-radiation equality is $\omega\, a_{\rm eq}=1/4$. The overall normalization of the scale factor is irrelevant to the final power-series solution for the fluid and metric variables in terms of $\tau$, and this convenient choice simplifies the equations.

To obtain a correct power-series for the super-horizon initial conditions, we conduct a normal mode analysis.
We begin with synchronous-gauge fluid and metric variables \citep{Ma1995, Bucher2000,Shaw2010magn} and equations of motion. We then define $x=k\eta$ and new fluid variables, dividing out the relative factors of $x$. 
That is, we set 
\bsub
\begin{align}
\tilde{\delta}_{i}=&~\delta_{i}/x,\\
t_{i}=&~\theta_{i}/x^{2},\\
\tilde{\sigma}_{\nu}=&~\sigma_{\nu}/x,\\
\tilde{F}_{\nu}^{\left(3\right)}=&~F_{\nu}^{\left(3\right)}/x^{2},
\end{align} 
\esub
where $\delta_i$,  $\theta_i$, $\sigma_\nu$ and $F^{(3)}_{\nu}$ are the usual fluid variables \citep[e.g., see][]{Ma1995}.
We then form a vector of the synchronous-gauge fluid and metric variables, ${U}^{\rm T}_{\vek{k}}=\left(\tilde{\delta}_{\gamma},\tilde{\delta}_{\nu},\tilde{\delta}_{\rm c},\tilde{\delta}_{\rm b},\tilde{t}_{\gamma \rm b},\tilde{t}_{\nu},\tilde{t}_{\rm c},\tilde{\sigma}_{\nu},\tilde{F}_{\nu}^{3},\Theta,\eta_{\rm m}\right)$, where for this analysis we work at times early enough that the tightly coupled photon-baryon fluid has a single velocity. In terms of the synchronous-gauge metric variable $h_{\rm m}$, we have $\Theta=h_{\rm m}'$, where $'$ denotes a derivative with respect to $x$. 
The full system of ODEs may then be written 
\begin{equation}
\frac{\id {U}_{\vek{k}}}{\id \ln x}
=\left(\underline{A}_{0}+\underline{A}_{1}x+ ... +\underline{A}_{n}x^{n}\right){U}_{k}\label{sa1},
\end{equation}
where $\underline{A}_{i}$ are matrices containing coefficients of terms of different order in $x$; the matrices are obtained by Taylor-expanding the conformal Hubble parameter $\mathcal{H}=\dot{a}/a=\omega \left(2\tau+1\right)/\left[\tau\left(\tau+1\right)\right]$, all homogeneous densities (baryons, CDM, neutrinos and photons), and pressures in powers of {$\tau\ll 1$}.  The space of solutions is spanned (to lowest order) by the eigenvectors ${U}_{\vek{k}}^{\lambda}$ (with eigenvalue $\lambda$) of $\underline{A}_{0}$:\begin{equation}
{U}_{\vek{k}}(\tau)=\sum_{\lambda} c_{\lambda} x^{\lambda}{U}_{\vek{k}}^{\left(\lambda\right)}.\label{sa2}
\end{equation} Here, $c_{\lambda}$ are coefficients setting the contribution of each normal mode to the solution, and can be chosen so that fluid variables match initial conditions. The physical growing normal modes are \footnote{Modes with $\lambda<0$ may still be `growing' modes, if the \textit{physical} variables $\delta,\theta,\sigma,...\propto x^{\gamma}$ for $\gamma>0$.}
\bsub
\begin{itemize}
\item{Adiabatic mode (AD): $\lambda=1$,
\begin{align}
U_{\vek{k}}^{\left(1\right)}&=\left(-\frac{1}{3}, -\frac{1}{3}, -\frac{1}{4}, -\frac{1}{4}, -\frac{1}{36},
-\frac{\left(23+4R_{\nu}\right)}{36 (15 + 4 R_{\nu})}, \right.
\nonumber\\
&\qquad\qquad\qquad\left.  
0,\frac{2}{3 (15 + 4 R_{\nu})}, 
\frac{4}{21 (15 + 4 R_{\nu})}, 1, 0
\right)^{\rm T},
\end{align}}
\item{Baryon isocurvature (BI) mode: $\lambda=-1$,
\begin{align}
U_{\vek{k}}^{(-1),~{\rm BI}}=\left(0, 0, 0, 1, 0, 0, 0, 0, 0, 0, 0\right)^{\rm T},
\end{align}}
\item{CDM isocurvature (CI) mode: $\lambda=-1$,
\begin{align}
U_{\vek{k}}^{(-1),~{\rm CI}}=\left(0, 0, 1, 0, 0, 0, 0, 0, 0, 0, 0\right)^{\rm T},
\end{align}}
\item{Neutrino density isocurvature (NDI) modes: $\lambda=-1$,
\begin{align}
U_{\vek{k}}^{(-1),~{\rm NDI} }=\left(-R_{\nu}/R_{\gamma}, 1, 0, 0, -R_{\nu}/(4R_{\gamma}), 1/4, 0, 0, 0, 0\right)^{\rm T},
\end{align}}
\item{Neutrino velocity isocurvature (NVI) mode, $\lambda=-2$,
\begin{align}
U_{\vek{k}}^{(-2)}=\left(0, 0, 0, 0, -R_{\nu}/R_{\gamma}, 1, 0, 0, 0, 0, 0\right)^{\rm T}.
\end{align}}
\end{itemize}
\esub

Around each normal mode, we can extend to a solution $\mathcal{U}_{\vek{k}}^{\lambda}(\tau)$ that includes higher order corrections: 
\begin{equation}
\mathcal{U}_{\vek{k}}^{(\lambda)}(\tau)=U_{\vek{k}}^{(\lambda)}x^{\lambda}+U_{\vek{k},\left(1\right)}^{(\lambda)}x^{\lambda+1}+...+U^{(\lambda)}_{\vek{k},\left(i\right)}x^{\lambda+i}+...\label{sa3},
\end{equation}
where the label ${U}_{\vek{k},\left(i\right)}^{(\lambda)}$ denotes the $i^{\rm th}$-order correction. 
We derive the corrections to the lowest-order solution by applying Eq.~(\ref{sa1}) to the \textit{ansatz}, Eq~(\ref{sa3}), obtaining \citep{Doran2003}:
\bsub
\begin{align}
\left[\left(\lambda+1\right)\mathcal{I}-\underline{A}_{0}\right]{U}^{(\lambda)}_{\vek{k},\left(1\right)}=&\underline{A}_{1}{U}^{(\lambda)}_{\vek{k}}\label{u1eqa},
\\
\left[\left(\lambda+2\right)\mathcal{I}-\underline{A}_{0}\right]{U}^{(\lambda)}_{\vek{k},\left(2\right)}=&\underline{A}_{1}{U}^{(\lambda)}_{\kappa,\left(1\right)}+\underline{A}_{2}{U}^{(\lambda)}_{\vek{k}},\label{u2eqa}
\\
\left[\left(\lambda+3\right)\mathcal{I}-\underline{A}_{0}\right]{U}^{(\lambda)}_{\vek{k},\left(3\right)}=&\underline{A}_{1}{U}^{(\lambda)}_{\vek{k},\left(2\right)}+\underline{A}_{2}{U}^{(\lambda)}_{\vek{k},\left(1\right)}+\underline{A}_{3}{U}^{(\lambda)}_{\vek{k},\left(1\right)},\label{u3eqa}
\\
\left[\left(\lambda+4\right)\mathcal{I}-\underline{A}_{0}\right]{U}_{\vek{k},\left(4\right)}^{(\lambda)}=&\underline{A}_{1}{U}^{(\lambda)}_{\vek{k},\left(3\right)}+\underline{A}_{2}{U}^{(\lambda)}_{\vek{k},\left(2\right)}
\nonumber\\
&\qquad
+\underline{A}_{3}{U}^{(\lambda)}_{\vek{k},\left(1\right)}+\underline{A}_{4}{U}^{(\lambda)}_{\vek{k}}.\label{u4eqa}
\end{align}
\esub
Here, $\mathcal{I}$ is the identity matrix in the space of all fluid+metric variables. The solutions to this linear system can yield higher order corrections to the time-evolution of the fluid variables for each normal mode.
For density isocurvature modes, the linear system Eq.~(\ref{u2eqa}) becomes under-constrained. By directly applying 
\begin{align}
k^{2}\eta_{\rm m}-\frac{\mathcal{H}}{2} \dot{h}_{\rm m}
=-4\pi G a^{2}\delta \rho
\label{einstein_constraint},
\end{align} 
the Einstein constraint equation, however, we may close the system to obtain $U_{\vek{k},\left(2\right)}^{(-1)}$ and continue to higher orders using Eqs. (\ref{u3eqa})-(\ref{u4eqa}). We thus reproduce the power series for the adiabatic and isocurvature modes of \citet{Bucher2000} and \citet{Shaw2010magn}. 

As a final step, we must perform a gauge transformation to CN gauge, in order to have initial conditions for \textsc{CosmoTherm}. The gauge transform is given by
\bsub
\begin{align}
\delta^{\rm con}_{i}=&~\delta^{\rm s}_{i}
+\alpha\,\frac{\dot{\overline{\rho}}_{i}}{\overline{\rho}_{i}},\\
\theta^{\rm con}_{i}=&~\theta^{\rm s}_{i}+\alpha \,k^{2},\\
\delta P^{\rm con}_{i}=&~\delta P^{\rm s}_{i}+\alpha\,{\dot{\overline{P}}},\\
\sigma^{\rm con}_{i}=&~\sigma_{i}^{\rm s},\\
\alpha(\vek{k},\tau)=&\frac{\dot{h}_{\rm m}+6\dot{\eta}_{\rm m}}{2k^{2}},\\
\psi=&
\frac{1}{2k^{2}}\left\{\ddot{h}_{\rm m}+6\ddot{\eta}_{\rm m}+2k^{2}\mathcal{H}\,\alpha\right \},
\\
\phi=& \eta_{\rm m}-\mathcal{H}\, \alpha,
\end{align}
\esub
where superscripts `${\rm s}$' denote synchronous gauge variables, `${\rm con}$' denote CN-gauge variables, $\delta P_{i}$ is the pressure perturbation of the $i^{\rm th}$ species and $\sigma_{i}$ the anisotropic stress in the $i^{\rm th}$ species. The homogeneous density/pressure of the $i^{\rm th}$ species are marked $\overline{\rho}_{i}$ and $\overline{P}_{i}$, while dots denote derivatives with respect to conformal time. The results of the gauge transformation are listed below.

{\sc CosmoTherm} evolves temperature variables for the photons and neutrinos, $\Theta_\ell$, so that for example $\delta_\gamma=4\Theta_{\gamma, 0}$, $\theta_\gamma=3k\Theta_1$ and $\sigma_\gamma=2\Theta_{\gamma, 2}$. The stiff ODE solver of {\sc CosmoTherm} can readily compute their evolution correctly, even if the only the leading order terms for the initial conditions are included. That is, starting the calculation at very early times, well within the super-horizon regime, terms of order $(k\eta)^2$ in the initial conditions can in principle be neglected without significantly affecting the solutions.

\subsection{Adiabatic mode (AD)}
The initial conditions for the well-known adiabatic mode are
\bsub
\beal
\psi&=\frac{10}{(15+4R_\nu)}=\frac{2}{3 a_\nu}
\\
\phi&=\frac{(10+4R_\nu)}{(15+4R_\nu)}=b_\nu\,\psi
\\
\delta_\gamma&=\delta_\nu=-\frac{20}{(15+4R_\nu)}=-2\psi
\\
\delta_{\rm c}&=\delta_{\rm b}=\frac{3}{4}\delta_\gamma=3\Theta_{\gamma, 0}
\\
\theta_{\gamma}&=\theta_{\nu}=\theta_{\rm b}=\theta_{\rm c}=\frac{1}{2}\,\psi\,k^2 \eta
\\
\sigma_{\nu}&=\frac{2\,(k \eta)^2}{3\left(15+4R_\nu\right)}=\frac{1}{15}\,\psi\,(k \eta)^2
=\frac{2}{15}\,\theta_{\nu}\,\eta.
\end{align}
\esub
It directly follows that the total initial entropy perturbation vanishes:
\beal
S(0, k)&=
R_{\rm c}\delta_{\rm c}+R_{\rm b}\delta_{\rm b}
-(3/4)(R_{\gamma}\delta_{\gamma}+R_{\nu}\delta_{\nu})
\nonumber\\
&=\delta_{\rm c}+R_{\rm b}(\delta_{\rm b}-\delta_{\rm c})-(3/4)[\delta_{\gamma}+R_{\nu}(\delta_{\nu}-\delta_\gamma)]\simeq 0
\end{align}
Thus this is an adiabatic (isentropic) initial condition, and initially only curvature perturbations are present.
A curvature perturbations of amplitude $\zeta(0, k)$ causes a potential perturbation $\psi=2\zeta(0, k)/(3 a_\nu)$.
Therefore, adiabatic modes enter the horizon with WKB amplitude $A\simeq -(3/2) \psi(0, k)$ \citep{Hu1996anasmall}. 

\subsection{Baryon and CDM isocurvature modes (BI/CI)}
For baryon isocurvature modes we have
\bsub
\beal
\psi&=-\frac{R_{\rm b}\tau}{2}\,\frac{\left(15-4R_\nu\right)}{\left(15+2R_\nu\right)} 
\\
\phi&=- \frac{R_{\rm b}\tau}{2}\,\frac{\left(15+4R_\nu\right)}{\left(15+2R_\nu\right)} 
=\frac{a_\nu}{c_\nu}\psi
\\
\delta_\gamma&=\delta_\nu=- 2R_{\rm b}\tau\,\frac{\left(15+4R_\nu\right)}{\left(15+2R_\nu\right)} =4\phi
\\
\delta_{\rm c}&=\delta_{\rm b}-1=- \frac{3}{2}\,R_{\rm b}\tau\,\frac{\left(15+4R_\nu\right)}{\left(15+2R_\nu\right)}=3\phi
\\
\theta_{\gamma}&=\theta_{\nu}=\theta_{\rm b}=-\frac{15}{2}\,R_{\rm b}\tau \,\frac{k^2\eta}{\left(15+2R_\nu\right)}
=\frac{\psi}{c_\nu}\,k^2 \eta
\\
\theta_{\rm c}&=-\frac{R_{\rm b}\tau}{6}\,\frac{\left(15-4R_\nu\right)}{\left(15+2R_\nu\right)}\,k^2 \eta=\frac{1}{3}\psi\,k^2 \eta
\\
\sigma_{\nu}&=-\frac{2\,R_{\rm b}\tau\,(k \eta)^2}{3\left(15+2R_\nu\right)}=\frac{4\psi}{45c_\nu}\,(k \eta)^2
=\frac{4}{45}\,\theta_\nu\,\eta.\label{bci_lsol}
\end{align}
\esub
The initial conditions for CDM isocurvature modes can be obtained from these expressions by replacing $R_{\rm b}\rightarrow R_{\rm c}$ and setting 
\beal
\delta_{\rm b}&=\delta_{\rm c}-1=- \frac{3}{2}\,R_{\rm c}\tau\,\frac{\left(15+4R_\nu\right)}{\left(15+2R_\nu\right)}=3\phi.
\end{align}
The baryon and CDM isocurvature modes both have vanishing initial potential perturbations (isocurvature condition) but total entropy perturbation $S(0, k)\simeq R_{\rm i}+\mathcal{O}(\tau)$ for $\rm i \in\{b, c\}$. 
This means that a baryon/CDM density perturbation with amplitude $\delta_{\rm i}(0, k)$ leads to an entropy perturbation $S_{\rm i}(0, k)=(\Omega_{\rm i}/\Omega_{\rm m})\,\delta_{\rm i}(0, k)$ initially. We also compute the gauge-invariant Bardeen curvature variable, $\zeta$, using the relation \citep{Shaw2010magn}
\begin{equation}
\zeta=\phi+2\frac{\left(\psi+\dot{\phi}/\mathcal{H}\right)}{3\left(1+w\right)},\end{equation}where $w$ is the total cosmic equation of state (equal to $1/3$ during radiation domination). Using Eqs.~(\ref{bci_lsol}), it is easy to see that $\zeta=0$ for the BI/CI modes when $\tau=\eta=0$, as should be the case for isocurvature perturbations.
These initial conditions only excite the $\sin(k\rs)$ term of the photon monopole transfer function. We therefore have $A\simeq 0$ and $B\simeq-\sqrt{6}/4(\Omega_{\rm i}/\Omega_{\rm m})(k_{\rm eq}/k)(1-4R_\nu/15)\,\delta_{\rm i}(0, k)$, where again $R_\nu$ accounts for the effect of anisotropic stress \citep{Hu1996anasmall}.

\subsection{Neutrino density isocurvature mode (NDI)}
Although these modes are called isocurvature modes, they do not have vanishing initial potential perturbations:
\bsub
\beal
\psi&=-\frac{2 R_{\nu}}{(15+4R_\nu)} =-\frac{2 R_{\nu}}{15a_\nu}
\\
\phi&=\frac{R_{\nu}}{(15+4R_\nu)} =- \frac{1}{2}\psi
\\
\delta_\gamma&=-\frac{R_\nu(11+8R_\nu)}{R_\gamma(15+4R_\nu)} 
=-\frac{R_\nu}{R_\gamma}+\phi
\\
\delta_\nu&=\frac{(15+8R_\nu)}{(15+4R_\nu)} =1+\phi
\\
\delta_{\rm c}&=\delta_{\rm b}=\frac{3 R_{\nu}}{(15+4R_\nu)}=3\phi
\\
\theta_{\gamma}&=\theta_{\rm b}=-\frac{19\,R_{\nu}\,k^2\eta}{4R_\gamma(15+4R_\nu)}
=\frac{19}{8R_\gamma}\, \psi\,k^2 \eta=\frac{\theta_{\nu}}{R_\gamma}+\frac{1}{2R_\gamma}\, \psi\,k^2 \eta
\\
\theta_{\nu}&=\frac{15\,k^2\eta}{4(15+4R_\nu)}=\frac{15}{8}\, \psi\,k^2 \eta
\\
\theta_{\rm c}&=-\frac{R_{\nu}}{(15+4R_\nu)}\,k^2 \eta=\frac{1}{2}\psi\,k^2 \eta
\\
\sigma_{\nu}&=\frac{(k \eta)^2}{2(15+4R_\nu)}=-\frac{\psi}{4R_\nu}\,(k \eta)^2
=\frac{\phi}{2R_\nu}\,(k \eta)^2=\frac{2}{15}\,\theta_{\nu}\,\eta.
\end{align}
\esub
An initial neutrino density perturbation of $\delta_\nu(0, k)$ leads to entropy perturbation $S(0, k)\simeq 4\,\phi(0, k) \simeq R_\nu\delta_\nu(0, k)/(15+4R_\nu)$ and potential perturbation $\psi(0, k) \simeq-2R_\nu\delta_\nu(0, k)/(15 a_\nu)$. Correspondingly, these modes excite both $\sin(k\rs)$ and $\cos(k\rs)$ parts of the photon monopole transfer function. As for the BI/CI modes, $\zeta=0$ initially for the NDI mode.

\subsection{Neutrino velocity isocurvature mode (NVI)}
The second isocurvature mode for neutrinos is sourced by non-vanishing velocity perturbations relative to the photons:
\bsub
\beal
\psi&=-\frac{4 R_{\nu}}{(5+4R_\nu)\,k\eta} 
\\
\phi&=\frac{4 R_{\nu}}{(5+4R_\nu)\,k\eta}  =- \psi
\\
\delta_\gamma&=\delta_\nu=\frac{16}{(5+4R_\nu)\,k\eta} =4\phi
\\
\delta_{\rm c}&=\delta_{\rm b}=\frac{12}{(5+4R_\nu)\,k\eta}=3\phi
\\
\theta_{\gamma}&=\theta_{\rm b}=
-\frac{R_\nu}{R_\gamma}\,k+\psi k^2 \eta
=-\frac{9\,R_{\nu}\,k}{R_\gamma(5+4R_\nu)}
=\frac{9}{4R_\gamma}\, \psi\,k^2 \eta
\\
\theta_{\nu}&=\frac{5k}{(5+4R_\nu)}=k+\psi\,k^2 \eta
\\
\theta_{\rm c}&=-\frac{4R_\nu k}{(5+4R_\nu)}=\psi\,k^2 \eta
\\
\sigma_{\nu}&=\frac{4 k\eta}{3(5+4R_\nu)}=-\frac{\psi}{3R_\nu}\,(k \eta)^2
=\frac{\phi}{3R_\nu}\,(k \eta)^2
\end{align}
\esub
These modes are very hard to excite and also show a coordinate divergence for small $k\eta$ \citep{Bucher2000}. Numerically, it is straightforward to integrate the corresponding perturbation equations with {\sc CosmoTherm}. Also, all physical quantities are gauge-independent and we find these to converge very well. As for the NDI mode, $\zeta=0$ initially for the NVI mode.

\end{appendix}

\bibliographystyle{mn2e}
\bibliography{Lit}

\end{document}